\documentclass[journal=jpcafh,manuscript=article]{achemso}

\usepackage[utf8]{inputenc}
\usepackage{hyperref}
\usepackage{textcomp}
\usepackage{mathptmx}
\usepackage{amsmath}
\usepackage{amssymb}
\usepackage{tikz}
\usepackage{graphicx}
\usepackage{epstopdf}
\usepackage[version=4]{mhchem}
\usepackage{multirow}
\usepackage[T1]{fontenc}
\usepackage{pgfplots}
\pgfplotsset{compat=1.18}
\usepgfplotslibrary{colorbrewer}

\pgfplotsset{cycle list/Set1-5, 
  cycle multiindex* list={
    linestyles \nextlist
    mark list* \nextlist
    Set1-5 \nextlist
    },
}
\AppendGraphicsExtensions{.tga}

\title{$G_0W_0$ Ionization Potentials of First-Row Transition Metal Aqua Ions}

\author{Daniel Mejia-Rodriguez}
\affiliation{Environmental Molecular Sciences Laboratory, Pacific Northwest National Laboratory, Richland, WA 99352, USA}
\alsoaffiliation{Physical and Computational Sciences Directorate, Pacific Northwest National Laboratory, Richland, WA 99352, USA}
\email{daniel.mejia@pnnl.gov}

\author{Alexander A. Kunitsa}
\affiliation{Zapata Computing, Inc., 100 Federal Street, Boston, MA 02110, USA}

\author{John L. Fulton}
\affiliation{Physical and Computational Sciences Directorate, Pacific Northwest National Laboratory, Richland, WA 99352, USA}

\author{Edoardo Apr\`a}
\affiliation{Environmental Molecular Sciences Laboratory, Pacific Northwest National Laboratory, Richland, WA 99352, USA}
\email{edoardo.apra@pnnl.gov}

\author{Niranjan Govind}
\affiliation{Physical and Computational Sciences Directorate, Pacific Northwest National Laboratory, Richland, WA 99352, USA}
\alsoaffiliation{Department of Chemistry, University of Washington, Seattle, WA 98195, USA}
\email{niri.govind@pnnl.gov}

\date{June 2023}

\begin{document}

\maketitle

\begin{abstract}

We report computations of the vertical ionization potentials within
the $GW$ approximation of the near-complete series of first-row
transition metal (V-Cu) aqua ions in their most common oxidation
states, i.e. V$^{3+}$, Cr$^{3+}$, Cr$^{2+}$, Mn$^{2+}$, Fe$^{3+}$,
Fe$^{2+}$, Co$^{2+}$, Ni$^{2+}$, and Cu$^{2+}$. The
\textit{d}-orbital occupancy of these systems spans a broad range from
\textit{d}$^{2}$ to \textit{d}$^{9}$. All the structures were first
optimized at the density functional theory level using a large cluster
of explicit water molecules that are embedded in a 
continuum solvation model. Vertical
ionization potentials were computed with the one-shot $G_0W_0$
approach on a range of transition metal ion clusters (6, 18, 40, and
60 explicit water molecules) wherein the convergence with respect to the
basis set size was evaluated using the systems with 40 water
molecules. We assess the results using three different density functional
approximations as starting points for the vertical ionization
potential calculations, namely $G_0W_0$@PBE, $G_0W_0$@PBE0, and
$G_0W_0$@r$^2$SCAN. While the predicted ground-state structures are
similar with all three exchange-correlation functionals, the vertical
ionization potentials were in closer agreement with experiment
when using the $G_0W_0$@PBE0 and
$G_0W_0$@r$^2$SCAN approaches, with the r$^2$SCAN based calculations
being significantly less expensive. Computed bond distances and vertical
ionization potentials for all structures were compared with available
experimental data and are in good agreement.
\end{abstract}

\section{Introduction}
Hydrated transition metal (TM) ions are involved in many biological
processes. In fact, five of the first-row TMs (Mn, Fe, Co, Cu, Zn) are essential to human health, while the other three (Cr, V, Ni) have
shown both beneficial and detrimental biological effects \cite{Crans:2020:104}. In many instances, TM aqua ions have very similar electronic structures to the ones in the active site of metalloenzymes. To that end, understanding their chemistry is important for the study of both biological systems and other
chemical systems with relevant industrial applications.  \cite{Stumm_1992,Richens:1997}.

The chemistry of TM aqua ions has been a very active
research topic for both experiment and theory. From a computational
point of view, the accurate description of these systems is still 
challenging despite many recent advances in the field \cite{Bhattacharjee:2022:1619}.
It is well known that a sufficiently accurate theoretical description of 3$d$ TM aqua ions 
must include, at least, the first and second solvation shells explicitly \cite{Yepes:2014:6850, Bhattacharjee:2022:1619}. An implicit 
solvation model, like the conductor-like screening model (COSMO) \cite{klamt1993},
must also be included in order to simulate bulk water effects and 
appropriately screen the highly-charged metal center. Very recently, Ghosh and co-workers have demonstrated that near-quantitative agreement between experimental
X-ray absorption spectroscopy (XAS) and theoretical simulations
can be achieved only by including a realistic description of the solvent and an appropriate level of theory. \cite{TM-XANES,Ghosh:2023:5203}
The resulting models were composed of 40 explicit water molecules around the charged TM center, described with a basis set of triple-$\zeta$ quality, surrounded by an implicit solvation description of the bulk solvent environment. 

In this manuscript, we have performed a comprehensive study of the
vertical ionization potentials (IPs) of the near-complete series of
first-row transition metal aqua ions (\textit{d}-orbital occupancy
from \textit{d}$^{2}$ to \textit{d}$^{9}$) within the $GW$
approximation\cite{Hedin:1965:A796} using our recently reported
scalable implementation based on Gaussian basis
sets.\cite{Mejia-Rodriguez:2021:7504,Mejia-Rodriguez:2022:4919}
To the best of our knowledge, this study is the first application of the {\color{blue} molecular} $GW$ method to the study of IPs of TM aqua ions. 
The $GW$ approximation is a well-established method in the solid-state physics community, \cite{Aryasetiawan:1998:237,Onida:2002:601,Martin:2016,Reining:2018:e1344} that is recently showing an increased interest in the chemistry arena for molecular systems\cite{Rostgaard:2010:085103,Blase:2011:115103,Bruneval2012,Caruso:2012:081102,VanSetten2013,Bruneval2016,vanSetten:2015:5665,Golze:2019:377,Golze:2020:1840,Mejia-Rodriguez:2021:7504}
due to a greatly improved balance between cost and accuracy. In particular, we have
utilized the one-shot $G_0W_0$ approach due to its lower computational
cost. Performing these calculations with other electronic structure
methods of similar accuracy would require a high{\color{blue}er} computational cost,
even when using linear-scaling approaches like the domain-based pair
natural orbital implementation of coupled cluster
theory.\cite{Bhattacharjee:2022:1619}

The rest of this paper is organized as follows. For completeness, we first present a brief overview of the theory. This is followed by the computational details, where we discuss the details of the calculations. The results and discussion are next followed by the conclusion.

\section{Theory}
\subsection{Overview of the \texorpdfstring{$GW$}{GW} Approximation}

%\subsection{Overview of the \texorpdfstring{$GW$}{GW} Approximation}

The $GW$ approximation (GWA)\cite{Hedin:1965:A796} is founded on the one-particle Green's function $G$, which describes the particle and hole scattering in the interacting many-body system, and the dynamically screened Coulomb interaction $W$ within linear response theory. To obtain the electron addition and removal energies, a non-local and dynamic effective potential, the self-energy $\Sigma$, expressed as a product of $G$ and $W$ is introduced. 
The self-energy $\Sigma$
replaces the mean-field exchange-correlation operator, and the $GW$
quasiparticle (QP) energies $\varepsilon_n^{GW}$ can be obtained as
corrections to the mean-field energies $\varepsilon_n$:
\begin{equation}
    \varepsilon_{n\sigma}^{GW} = \varepsilon_{n\sigma} + \Re\left( \Sigma_{n\sigma}(\varepsilon_{n\sigma}^{GW}) \right) - V_{n\sigma}^{xc}
    \label{eq:qpeq}
\end{equation}
where $\sigma$ is the spin index, and $V_{n\sigma}^{xc}$ and $\Sigma_{n\sigma}$ denote the $n$th diagonal element of the corresponding matrix representation in the molecular orbital basis. Equation \ref{eq:qpeq} is non-linear and must be solved iteratively.

The self-energy operator $\Sigma$ is given in terms of the Green's function $G^{\sigma}$ and the screened Coulomb interaction $W$ as
\begin{equation}
    \Sigma_\sigma(\mathbf{r},\mathbf{r}',\omega) = \frac{i}{2\pi} \int \mathrm{d}\xi G^{\sigma}(\mathbf{r},\mathbf{r}',\omega+\xi) W(\mathbf{r},\mathbf{r}',\xi) e^{i\xi\eta}
\end{equation}
with $\eta$ being a positive infinitesimal. In practice, however, the GWA is often performed as a one-shot perturbative approach commonly known as $G_0W_0$. Here, $G_0^{\sigma}$ is the non-interacting mean-field Green's function,
\begin{equation}
    G_0^\sigma(\mathbf{r},\mathbf{r}',\omega) = \sum\limits_m \frac{\phi_{m\sigma}(\mathbf{r})\phi_{m\sigma}(\mathbf{r}')}{\omega - \varepsilon_{m\sigma} - i\eta \; \mathrm{sign}(\mu - \varepsilon_{m\sigma})} \;\; ,
\end{equation}
and $W_0$ is obtained using the random phase approximation (RPA) as
\begin{equation}
    W_0(\mathbf{r},\mathbf{r}',\omega) = \int \mathrm{d}\mathbf{r}'' \epsilon^{-1} (\mathbf{r},\mathbf{r}'',\omega) v(\mathbf{r}'',\mathbf{r}')
\end{equation}
In the preceding equations, $\mu$ is the Fermi-level of the system, $v(\mathbf{r},\mathbf{r}')$ is the bare Coulomb interaction, and $\epsilon(\mathbf{r},\mathbf{r}'',\omega)$ is the RPA dynamical dielectric function:
\begin{equation}
    \epsilon(\mathbf{r},\mathbf{r}'',\omega) = \delta(\mathbf{r},\mathbf{r}') - \int \mathrm{d}\mathbf{r}'' v(\mathbf{r},\mathbf{r}'')\chi_0(\mathbf{r}'',\mathbf{r}',\omega)
\end{equation}
In the RPA, the irreducible polarizability $\chi_0$ has a sum-over-states representation\cite{Adler:1962,Wiser:1963}
\begin{equation}
    \chi_0(\mathbf{r},\mathbf{r}',\omega) = \sum\limits_\sigma \sum\limits_{i,a}  \left[ \frac{\phi_{i\sigma}(\mathbf{r})\phi_{a\sigma}(\mathbf{r})\phi_{i\sigma}(\mathbf{r}')\phi_{a\sigma}(\mathbf{r}')}{\omega - \varepsilon_{a\sigma} + \varepsilon_{i\sigma} +i\eta} + \frac{\phi_{i\sigma}(\mathbf{r})\phi_{a\sigma}(\mathbf{r})\phi_{i\sigma}(\mathbf{r}')\phi_{a\sigma}(\mathbf{r}')}{-\omega - \varepsilon_{a\sigma} + \varepsilon_{i\sigma} + i\eta} \right]
\end{equation}
where the index $i$ runs over the occupied orbitals while the index $a$ runs over the virtual ones. Since a detailed discussion of the theory is beyond the scope of this paper, we refer the reader to comprehensive reviews on the subject.\cite{Onida:2002:601,Martin:2016,Reining:2018:e1344}

We have recently implemented the GWA for molecular systems in the open-source NWChem computational chemistry program within the Gaussian basis set framework. Both spectral decomposition (SD) and contour-formation (CD) methods were implemented and demonstrated to perform well for valence and core ionization spectra. Here, we utilize the CD-$GW$ technique for the integration of $\Sigma$. For further details of our implementation and performance can be found in Refs.\citenum{Mejia-Rodriguez:2021:7504,Mejia-Rodriguez:2022:4919}

\iffalse
as it has been shown to have a good cost-effective profile for the evaluation of CLBEs. Further details about CD-$GW$ and its implementation using local orbitals can be obtained from References \nocite{Golze:2018:4856,Holzer:2019:204116,MejiaRodriguez:2021:7504} \citenum{Golze:2018:4856}, \citenum{Holzer:2019:204116}, and \citenum{Mejia-Rodriguez:2021:7504} for example.
\fi

\section{Computational Details}

In order to perform a $G_0W_0$ calculation, one needs to first compute the single-particle orbitals and energies obtained with a mean-field theory. Given that $G_0W_0$ does not optimize these orbitals, the method has some degree of starting point dependency. Here, we start from Kohn-Sham (KS) density functional theory (DFT) calculations using three different density functional approximations (DFAs) to the exchange-correlation energy {\color{blue} with empirical dispersion corrections}. All ground-state DFT calculations were performed with the latest development version of the NWChem\cite{NWChem} computational chemistry software using the Perdew-Burke-Ernzerhoff (PBE)\cite{pbe}, its global hybrid PBE0\cite{pbe0} extension, and the recent regularized-restored strongly constrained and appropriately normed (r$^2$SCAN)\cite{r2scan} DFAs. The def2-\{T,Q\}ZVP basis set\cite{def2} for hydrogen and oxygen were used in combination with the Sapporo-\{T,Q\}ZP basis sets for the TM atoms\cite{sapporo}, respectively. {\color{blue} The choice of the Sapporo basis set to describe the TM atoms was motivated by the explicit consideration of core-valence correlations in developing such set\cite{sapporo} while still retaining a more compact representation than other bases, such as the cc-pCV$n$Z and cc-pwCV$n$Z families \cite{ccpcvnz1,ccpcvnz2,ccpcvnz3}. Moreover, a similar basis set combination was recently used for the accurate description of the X-ray absorption spectroscopy of these systems \cite{TM-XANES}.}
The van der Waals dispersion interactions were accounted for with the D3 empirical model\cite{DFTD3} using Becke-Johnson damping\cite{BJdamping}. D3 parameters for r$^2$SCAN were obtained from Ref. \nocite{r2scand4}\citenum{r2scand4}.

The explicitly solvated TM aqua ion clusters were optimized at the corresponding DFT-level without symmetry constraints, surrounded by the COSMO implicit solvation model\cite{klamt1993} as modified by York and Karplus \cite{york-karplus}. Different from traditional implementations of the COSMO model, atomic spheres were discretized using a spherical Fibonacci lattice \cite{Dixon1989,Svergun1994,swinbank2006,Gonzalez2010} with 401 points. The York-Karplus switching function parameters, including the covalent radii, were updated due to the presence of many COSMO charges located inside the explicit water volume. Further details about these modifications can be found in the Supporting Information. The starting geometries for the optimization of the \ce{[M(H2O)6]^n+} and \ce{[M(H2O)18]^n+} aqua ions were taken from Ref. \nocite{Uudsemaa2003}\citenum{Uudsemaa2003}, the starting geometries for the larger \ce{[M(H2O)40]^n+} systems were taken from Ref.\citenum{TM-XANES}, and the starting geometries for the \ce{[M(H2O)60]^n+} were obtained by carving a sphere out from an equilibrated {\color{blue} quantum mechanics/molecular mechanics hybrid simulation using the} SPC/E {\color{blue} water model} and pre-optimized using a double-$\zeta$ basis set.

Vertical ionization energies were computed using the $GW$ module recently implemented in a development version of NWChem\cite{GWnwchem,DMRGWBasis} at the $G_0W_0$ level. {\color{blue} The ground-state KS orbitals were obtained in the presence of the implicit solvation model.} The necessary integrals were obtained using the contour-deformation approach using 200 points for the numerical integration over the imaginary axis. The def2-\{T,Q\}ZVP-RIFIT charge density fitting basis\cite{RIFIT,BSE} was used for H and O. The fitting basis for the TM atom was automatically generated using the ``AutoAux'' approach described in Ref. \nocite{Stoychev:2017:554}\citenum{Stoychev:2017:554} and recently implemented in the Basis Set Exchange\cite{BSE}.

Molecular dynamics (MD) simulations were performed with NWChem with the quantum molecular dynamics (QMD) module.\cite{fischer:2016:1429} These simulations were performed under the canonical ensemble using the Nos\'e-Hoover chain thermostat\cite{Nose:1984:511,Hoover:1985:1695,Martyna:1992:2635} with three oscillators, a target temperature of 298 K, and a coupling frequency of 300 cm$^{-1}$. Trajectories were collected for at least 3.5 ps using a time step of 0.4838 fs (20 a.u., 7235 steps), and the first 1 ps was left out from the analysis. All trajectories were generated using the def2-SVP double-$\zeta$ orbital basis set, the def2-universal-jfit fitting basis set\cite{JFIT,BSE}, and the PBE exchange-correlation functional.

\section{Results and Discussion}
\subsection{Vertical ionization potentials with optimized geometries}
The quality of the optimized geometries was assessed by comparing 
to the experimental results obtained
via fitting of the extended X-ray absorption fine structure (EXAFS) 
spectra \cite{Miyanaga:1988,Lundberg:2007,Fulton:2012:2588,Persson:2020:9538}
of the hydrated TM ions. Figure \ref{fig:bonds} shows
average TM-oxygen distances for the first solvation shell.  
The theoretical averages were computed using the six water molecules for
the \ce{V^3+}, \ce{Cr^3+}, \ce{Mn^2+}, \ce{Fe^3+}, \ce{Fe^2}, \ce{Co^2+}, and \ce{Ni^2+} {\color{blue} aqua} ions, or the subset of equatorial water molecules
for \ce{Cr^2+}, and \ce{Cu^2+} aqua ions. 
Note that we only focus on such bond averages because those are the quantities
available from the EXAFS fitting (see Supplementary
Information for individual bond distances {\color{blue} and Figure \ref{fig:bondsall} for fluctuations of such distances during an MD simulation}).
Clear trends can be observed 
upon increasing the size of the water cluster around the cations 
where most of the ion-water bonds elongate by about 0.03 \AA, while the
equatorial \ce{Cr^2+}--\ce{O} and \ce{Cu^2+}--\ce{O} bonds shorten by the
same amount. These bond distances are stable with respect to the change of
basis set, from triple-$\zeta$ to quadruple-$\zeta$, at least for
the smaller \ce{[M(H2O)6]^n+} and \ce{[M(H2O)18]^n+}
systems. The average bond distances with the PBE0 exchange-correlation functional, middle panel of Figure \ref{fig:bonds},
for \ce{[M(H2O)40]^n+} are in very good agreement with those obtained from 
Ref. \citenum{TM-XANES} using a different basis set for the water molecules, namely
6-311G**.\cite{krishnan1980a}
It is interesting to note that the
r$^2$SCAN-D3 functional yields geometries comparable with the PBE0-D3 global hybrid functional, but at a fraction of the
computational cost.\footnote{where the speed-up comes from the use of the density fitting approach for the evaluation of the Coulomb potential in the computation of the ground state energy.}
Overall, all three density functional approximations (DFAs) perform rather well in predicting the average
conformation of the first hydration shell, although PBE0-D3 and r$^2$SCAN-D3 
yield slightly better results than the ``pure'' PBE-D3 exchange-correlation functional.

\begin{figure}
    \centering
    \begin{tikzpicture}
    \begin{axis}[
    name=plot1,
    legend pos = outer north east,
    width=12cm, height=7cm,
    ylabel={$r_{\text{M-O}}$ [\AA]},
    y tick label style={/pgf/number format/.cd, fixed, fixed zerofill, precision=2, /tikz/.cd},
    xmin=0, xmax=10,
    ymin=1.90, ymax=2.25,
    scaled y ticks=false,
    xtick={1,2,3,4,5,6,7,8,9},
    xticklabel style = {align=center},
    xticklabels={,,}
    ]
    \addplot+[unbounded coords=jump, line width=2pt, mark options={scale=1.5}] table[x=X, y expr=\thisrow{PBE6W}] {bonds.dat};
    \addplot+[unbounded coords=jump, line width=2pt, mark options={scale=1.5}] table[x=X, y expr=\thisrow{PBE18W}] {bonds.dat};
    \addplot+[unbounded coords=jump, line width=2pt, mark options={scale=1.5}] table[x=X, y expr=\thisrow{PBE40W}] {bonds.dat};
    \addplot+[unbounded coords=jump, line width=2pt, mark=none, black, solid] table[x=X, y expr=\thisrow{EXPTL}] {bonds.dat};
    \draw (4.5,2.20) node [below right] {PBE};
    \legend{\ce{[M(H2O)6]^n+},\ce{[M(H2O)18]^n+},\ce{[M(H2O)40]^n+},Expt} 
    \end{axis}

    \begin{axis}[
    name=plot2,
    at=(plot1.below south east),
    anchor=above north east,
    legend pos = outer north east,
    width=12cm, height=7cm,
    ylabel={$r_{\text{M-O}}$ [\AA]},
    y tick label style={/pgf/number format/.cd, fixed, fixed zerofill, precision=2, /tikz/.cd},
    xmin=0, xmax=10,
    ymin=1.90, ymax=2.25,
    scaled y ticks=false,
    xtick={1,2,3,4,5,6,7,8,9},
    xticklabels={,,}
]
    \addplot+[unbounded coords=jump, line width=2pt, mark options={scale=1.5}] table[x=X, y expr=\thisrow{PBE06W}] {bonds.dat};
    \addplot+[unbounded coords=jump, line width=2pt, mark options={scale=1.5}] table[x=X, y expr=\thisrow{PBE018W}] {bonds.dat};
    \addplot+[unbounded coords=jump, line width=2pt, mark options={scale=1.5}] table[x=X, y expr=\thisrow{PBE040W}] {bonds.dat};
    \addplot+[unbounded coords=jump, line width=2pt, mark=none, black, solid] table[x=X, y expr=\thisrow{EXPTL}] {bonds.dat};
    \draw (4.5,2.20) node [below right] {PBE0};
    \legend{\ce{[M(H2O)6]^n+},\ce{[M(H2O)18]^n+},\ce{[M(H2O)40]^n+},Expt} 
    \end{axis}

    \begin{axis}[
    name=plot3,
    at=(plot2.below south east),
    anchor=above north east,
    legend pos = outer north east,
    width=12cm, height=7cm,
    ylabel={$r_{\text{M-O}}$ [\AA]},
    y tick label style={/pgf/number format/.cd, fixed, fixed zerofill, precision=2, /tikz/.cd},
    xmin=0, xmax=10,
    ymin=1.90, ymax=2.25,
    scaled y ticks=false,
    xtick={1,2,3,4,5,6,7,8,9},
    xticklabel style = {align=center},
    xticklabels={\ce{V^3+}\\$d^2$,\ce{Cr^3+}\\$d^3$,\ce{Cr^2+}\\$d^4$,\ce{Mn^2+}\\$d^5$,\ce{Fe^3+}\\$d^5$,\ce{Fe^2+}\\$d^6$,\ce{Co^2+}\\$d^7$,\ce{Ni^2+}\\$d^8$,\ce{Cu^2+}\\$d^9$}]
    \addplot+[unbounded coords=jump, line width=2pt, mark options={scale=1.5}] table[x=X, y expr=\thisrow{R2SCAN6W}] {bonds.dat};
    \addplot+[unbounded coords=jump, line width=2pt, mark options={scale=1.5}] table[x=X, y expr=\thisrow{R2SCAN18W}] {bonds.dat};
    \addplot+[unbounded coords=jump, line width=2pt, mark options={scale=1.5}] table[x=X, y expr=\thisrow{R2SCAN40W}] {bonds.dat};
    \addplot+[unbounded coords=jump, line width=2pt, mark options={scale=1.5}] table[x=X, y expr=\thisrow{R2SCAN60W}] {bonds.dat};
    \addplot+[unbounded coords=jump, line width=2pt, mark=none, black, solid] table[x=X, y expr=\thisrow{EXPTL}] {bonds.dat};
    \draw (4.5,2.20) node [below right] {r$^2$SCAN};
    \legend{\ce{[M(H2O)6]^n+},\ce{[M(H2O)18]^n+},\ce{[M(H2O)40]^n+},\ce{[M(H2O)60]^n+},Expt} 
    \end{axis}

    \end{tikzpicture}
    \caption{Average metal-oxygen bond distances, in \AA, obtained with three different exchange-correlation functionals. Complexes were modeled with 6, 18, 40, or 60 explicit water molecules surrounded by COSMO implicit solvation. The hydrated \ce{Cr^2+} and \ce{Cu^2+} average bond distances only took into account equatorial bonds.}
    \label{fig:bonds}
\end{figure}

\begin{figure}
    \centering
    \begin{tikzpicture}
    \begin{axis}[
    name=plot1,
    legend pos = outer north east,
    width=12cm, height=7cm,
    ylabel={IP\textsubscript{$G_0W_0$} [eV]},
    y tick label style={/pgf/number format/.cd, fixed, fixed zerofill, precision=2, /tikz/.cd},
    xmin=0, xmax=10,
    ymin=5.0, ymax=15.00,
    scaled y ticks=false,
    xtick={1,2,3,4,5,6,7,8,9},
    xticklabels={,,}
]
    \addplot+[unbounded coords=jump, line width=2pt, mark options={scale=1.5}] table[x=X, y expr=\thisrow{PBE6W}] {ips.dat};
    \addplot+[unbounded coords=jump, line width=2pt, mark options={scale=1.5}] table[x=X, y expr=\thisrow{PBE18W}] {ips.dat};
    \addplot+[unbounded coords=jump, line width=2pt, mark options={scale=1.5}] table[x=X, y expr=\thisrow{PBE40W}] {ips.dat};
    \addplot+[unbounded coords=jump, line width=2pt, mark=none, black, solid] table[x=X, y expr=\thisrow{EXPTL}] {ips.dat};
    \draw (6.5,13.0) node [below right] {PBE};
    \legend{\ce{[M(H2O)6]^n+},\ce{[M(H2O)18]^n+},\ce{[M(H2O)40]^n+},Expt}
    \end{axis}

    \begin{axis}[
    name=plot2,
    ylabel={IP\textsubscript{$G_0W_0$} [eV]},
    at=(plot1.below south east),
    anchor=above north east,
    legend pos = outer north east,
    width=12cm, height=7cm,
    xmin=0, xmax=10,
    ymin=5.00, ymax=15.00,
    y tick label style={/pgf/number format/.cd, fixed, fixed zerofill, precision=2, /tikz/.cd},
    scaled y ticks=false,
    xtick={1,2,3,4,5,6,7,8,9},
    xticklabels={,,}
]
    \addplot+[unbounded coords=jump, line width=2pt, mark options={scale=1.5}] table[x=X, y expr=\thisrow{PBE06W}] {ips.dat};
    \addplot+[unbounded coords=jump, line width=2pt, mark options={scale=1.5}] table[x=X, y expr=\thisrow{PBE018W}] {ips.dat};
    \addplot+[unbounded coords=jump, line width=2pt, mark options={scale=1.5}] table[x=X, y expr=\thisrow{PBE040W}] {ips.dat};
    \addplot+[unbounded coords=jump, line width=2pt, mark=none, black, solid] table[x=X, y expr=\thisrow{EXPTL}] {ips.dat};
    \draw (6.5,13.0) node [below right] {PBE0};
    \legend{\ce{[M(H2O)6]^n+},\ce{[M(H2O)18]^n+},\ce{[M(H2O)40]^n+},Expt} 
    \end{axis}

    \begin{axis}[
    name=plot3,
    ylabel={IP\textsubscript{$G_0W_0$} [eV]},
    at=(plot2.below south east),
    anchor=above north east,
    legend pos = outer north east,
    width=12cm, height=7cm,
    xmin=0, xmax=10,
    ymin=5.00, ymax=15.00,
    y tick label style={/pgf/number format/.cd, fixed, fixed zerofill, precision=2, /tikz/.cd},
    scaled y ticks=false,
    xtick={1,2,3,4,5,6,7,8,9},
    xticklabel style = {align=center},
    xticklabels={\ce{V^3+}\\$d^2$,\ce{Cr^3+}\\$d^3$,\ce{Cr^2+}\\$d^4$,\ce{Mn^2+}\\$d^5$,\ce{Fe^3+}\\$d^5$,\ce{Fe^2+}\\$d^6$,\ce{Co^2+}\\$d^7$,\ce{Ni^2+}\\$d^8$,\ce{Cu^2+}\\$d^9$}]
    \addplot+[unbounded coords=jump, line width=2pt, mark options={scale=1.5}] table[x=X, y expr=\thisrow{R2SCAN6W}] {ips.dat};
    \addplot+[unbounded coords=jump, line width=2pt, mark options={scale=1.5}] table[x=X, y expr=\thisrow{R2SCAN18W}] {ips.dat};
    \addplot+[unbounded coords=jump, line width=2pt, mark options={scale=1.5}] table[x=X, y expr=\thisrow{R2SCAN40W}] {ips.dat};
    \addplot+[unbounded coords=jump, line width=2pt, mark options={scale=1.5}] table[x=X, y expr=\thisrow{R2SCAN60W}] {ips.dat};
    \addplot+[unbounded coords=jump, line width=2pt, mark=none, black, solid] table[x=X, y expr=\thisrow{EXPTL}] {ips.dat};
    \draw (6.5,13.0) node [below right] {r$^2$SCAN};
    \legend{\ce{[M(H2O)6]^n+},\ce{[M(H2O)18]^n+},\ce{[M(H2O)40]^n+},\ce{[M(H2O)60]^n+},Expt} 
    \end{axis}

    \end{tikzpicture}
    \caption{$G_0W_0$ vertical IPs, in eV, using orbitals obtained with three different exchange-correlation functionals.}
    \label{fig:vip}
\end{figure}

Figure \ref{fig:vip} shows the
IPs obtained using the $G_0W_0$@DFT approach for the 
optimized cluster geometries of the hydrated hexaaqua complexes and experimental reference values obtained from
Ref. \citenum{Yepes:2014:6850}. The theoretical IPs were obtained using
the triple-$\zeta$ basis set combination (def2-TZVP/Sapporo-TZP-2012) and,
as a consequence, the values are not converged with respect
to the basis set size (see below). However,
the trends are clear: $G_0W_0$@PBE0 overestimates the IPs of all aqua
ions, while $G_0W_0$@PBE and $G_0W_0$@r$^2$SCAN underestimate them. 
Figure \ref{fig:vip} also shows that there are still some minimal variations
between the 18- and 40-explicit-water models. It is important to note
that our $G_0W_0$ implementation does not take into account the optical, {\color{blue} i.e fast,} response 
from the implicit solvation model. The effect of this omission is expected to be
small since we are focusing on the occupied part of the spectrum of
systems with relatively large charges\cite{Duchemin:2016:164106} {\color{blue} and the contribution from the static dielectric constant is already taken into account by the ground-state KS calculation}.

In order to further assess the convergence of the $G_0W_0$@DFT IP values
 with respect to the explicit solvation size,
and to evaluate indirectly the effect of the missing optical response term mentioned above,
we decided to extend the size of the explicit solvation shell to 60 water molecules
using the r$^2$SCAN DFA. The choice of DFA follows from the quality of
geometrical parameters and $G_0W_0$ vertical IPs that can be obtained 
with r$^2$SCAN at a
fraction of the cost of PBE0 calculations. These results are 
shown in the bottom panels of  Figures \ref{fig:bonds} and \ref{fig:vip}.
It is evident that the \ce{[M(H2O)40]^n+} systems already offer
converged results for both bond distances and IPs. This means that, as
expected, the missing response term should be negligible in the larger
models.

Although $G_0W_0$@PBE0 seems to offer the best match to the experimental values,
it is likely that larger basis sets and (partial) self-consistent $GW$ approaches 
will lead to a systematic overestimation of the IPs. In contrast, 
$G_0W_0$@r$^2$SCAN IPs extrapolated to the basis set limit still underestimate the
experimental results, albeit with smaller deviations (see below). It is
likely that the same trend will be found with $G_0W_0$@PBE IPs extrapolated to the 
basis set limit. Overall, $G_0W_0$@DFT predicts IPs of similar
 quality as other $\Delta$SCF methodologies, including $\Delta$DFT and $\Delta$CCSD(T)\cite{Yepes:2014:6850,Bhattacharjee:2022:1619} {\color{blue}(See Table S2 and Figure S3 and S4 in the Supplementary Information)}. The reasonably good
 IPs predicted by all these methods indicate that the multi-reference character of the
 hydrated 3$d$ TM ions is moderate to small, as expected for high-spin complexes 
 bound to ligands with weak field strengths\cite{Duan:2022:4962}.
 {\color{blue} Unfortunately, a direct comparison between our $G_0W_0$@DFT IPs and the CCSD(T) results of Ref. \citenum{Bhattacharjee:2022:1619} is not possible because the GWA does not give access to adiabatic IPs.
 
We would also like to point out that our $G_0W_0$ implementation has a lower computational cost than the $\Delta$DFT method for the small \ce{[M(H2O)6]^{n+}} clusters, and a comparable cost for the larger systems: assuming that the DFT calculation of the $n$ highest hole states can be converged as easily as the ground state, actual timings from the $G_0W_0$@r$^2$SCAN calculation of the \ce{[M(H2O)60]^{n+}} clusters suggest that $G_0W_0$ would be $8.2/n + 1.12$-times slower than the $\Delta$DFT approach (factor 2 slower for $n=9$). This assumption is not always valid, and the $8.2/n + 1.12$ factor is more likely to represent an upper bound due to convergence issues. The factor will be even smaller when the ground state KS calculation is performed with a hybrid functional because each KS iteration would be slower and the $G_0W_0$ quasiparticle equations are easier to converge with these orbitals. }

\subsection{Basis set convergence} \label{sec:basis}

\begin{figure}
\begin{tikzpicture}
    \begin{axis}[
    name=plot3,
    ylabel={IP\textsubscript{$G_0W_0$} [eV]},
    anchor=above north east,
    legend pos = north west,
    width=12cm, height=7cm,
    xmin=0, xmax=8,
    ymin=7.0, ymax=11.0,
    y tick label style={/pgf/number format/.cd, fixed, fixed zerofill, precision=2, /tikz/.cd},
    scaled y ticks=false,
    xtick={1,2,3,4,5,6,7},
    xticklabels={\ce{V^3+},\ce{Cr^3+},\ce{Mn^2+},\ce{Fe^3+},\ce{Co^2+},\ce{Ni^2+},\ce{Cu^2+}}]
    \addplot+[line width=2pt, mark options={scale=1.5}] table[x=X, y expr=\thisrow{R2SCAN40W}] {ips2.dat};
    \addplot+[line width=2pt, mark options={scale=1.5}] table[x=X, y expr=\thisrow{R2SCAN40WQZ}] {ips2.dat};
    \addplot+[line width=2pt, mark options={scale=1.5}] table[x=X, y expr=\thisrow{R2SCAN40WCBS}] {ips2.dat};
    \addplot+[line width=2pt, mark options={scale=1.5}] table[x=X, y expr=\thisrow{R2SCAN40WCBS2}] {ips2.dat};
    \addplot+[line width=2pt, mark=none, black, solid] table[x=X, y expr=\thisrow{EXPTL}] {ips2.dat};
    \draw[thin] (axis cs:\pgfkeysvalueof{/pgfplots/xmin},0) -- (axis cs:\pgfkeysvalueof{/pgfplots/xmax},0);
    \draw (5,0.5) node [below right] {\ce{[M(H2O)40]^n+}};
    \legend{TZ,QZ,$\zeta(4)$,$\zeta(3)$,Exptl.} 
    \end{axis}
    \end{tikzpicture}
    \caption{$G_0W_0$@r$^2$SCAN vertical IPs, in eV. The potentials were obtained with def2-TZVP/Sapporo-2012-TZP (TZ) or def2-QZVP/Sapporo-2012-QZP (QZ). Two 2-point complete basis set (CBS) extrapolations, based on the $\zeta(4)$ and $\zeta(3)$ Riemann formula (see text for details), are also shown.}
    \label{fig:basis}
\end{figure}

The convergence with respect to the basis set size was evaluated using a subset of the [M(H$_2$O)$_{40}$]$^{n+}$ clusters only with r$^2$SCAN densities as starting point. Figure \ref{fig:basis} shows that the use of a quadruple-$\zeta$ basis set combination (def2-QZVP/Sapporo-QZP-2012) leads to IPs about 0.30 eV larger than their triple-$\zeta$ (def2-TZVP/Sapporo-TZP-2012) counterparts for all TM aqua ions except for \ce{Fe^3+}. A two-point complete basis set (CBS) extrapolation, using the Riemann $\zeta$ function technique\cite{CBSRiemann}, 

\begin{equation}
    \label{eq:CBSRiemann}
    E_{\infty} = E_L + L^4\left(E_L - E_{L-1}\right)\left(\zeta(4) - \sum\limits_{l=1}^L l^{-4}\right)
\end{equation}
where $L$ is the cardinality of the largest basis set and the Riemann function $\zeta(4) = \pi^4/90$, adds another 0.27 eV to the IP value, on average. 
 It has been shown that the $\zeta(4)$-based CBS correction has the same leading term as the conventional formula $E_{L} = E_{\infty} + A \; L^{-3}$ for large $L$, but that the $\zeta(4)$-based approach offer significantly better extrapolations for correlation energies \cite{CBSRiemann}. However, Bruneval and co-workers \cite{Bruneval2016} have noted that $GW$ quasiparticle energies might have a better fit to the empirical $E_{L} = E_{\infty} + A \; L^{-2}$ formula. As a consequence, we have also adapted the Riemann approach to have the same $L^{-2}$ leading term:
\begin{equation}
    \label{eq:CBSRiemann2}
    E_{\infty} = E_L + L^3\left(E_L - E_{L-1}\right)\left(\zeta(3) - \sum\limits_{l=1}^L l^{-3}\right)
\end{equation}
where $\zeta(3) \approx 1.202057$. The $\zeta(3)$-based CBS extrapolation adds, on average, 0.45 eV to the IPs obtained with the quadruple-$\zeta$ basis set combination.

The $\zeta(4)$-extrapolated $G_0W_0$@r$^2$SCAN IPs underestimate the experimental ones by 0.30 eV on average, with the most significant discrepancy observed for \ce{Co^2+} (-0.50 eV deviation). The $\zeta(3)$-extrapolated IPs still underestimate the experimental reference by 0.13 eV on average. We note, however, that these results contain no dynamical information (see below) and that the more advanced, but computationally more expensive, ev$GW$ partially self-consistent method might add several tenths of eV to these results \cite{Hung:2017:2135}.
These results are comparable to the ones reported by Yepes and co-workers\cite{Yepes:2014:6850} using the $\Delta$SCF approach with global hybrid density functional approximations. However, we note that the computational setup used by Yepes and co-workers  may not be well converged due to the small cluster and basis set sizes, with the latter expected to have the largest influence.

\subsection{Structural sampling through \emph{ab initio} molecular dynamics simulations}
It is important to note that no dynamical information is contained in the previous results. That is, the bond distances and IPs were obtained from a single optimal conformation of each model.
It is well-known that Cu$^{2+}$ exhibits complex first-solvation shell dynamics \cite{Pasquarello:2001:856,Frank:2018:204302,Persson:2020:9538} with a rather fast solvent exchange rate.\cite{Helm:2005:1923,Lincoln:2005:523}
Other 3$d$ divalent cations have water exchange rate constants similar to those of Cu$^{2+}$ (see Table \ref{tab:rates}), and might also exhibit large fluctuations within their first solvation shell.
\begin{table}[]
    \centering
    \caption{Experimental water exchange rate constants [s$^{-1}$] for the first-row transition metal ions considered in this study. All values taken from Ref. \citenum{Helm:2005:1923} and \citenum{Lincoln:2005:523}.} \label{tab:rates}
    \begin{tabular}{c c c c}
    Metal         & $d$-Electron config. & $k_{\text{H}_2\text{O}}$ (298 K) \\\hline
     Cu$^{2+}$    &  $d^9$ &  $5.7\times 10^9$  \\
     Cr$^{2+}$    &  $d^4$ &  $>1.2\times 10^8$  \\
     Mn$^{2+}$    &  $d^5$ &  $2.1 \times 10^7$  \\
     Fe$^{2+}$    &  $d^6$ &  $4.4 \times 10^6$  \\
     Co$^{2+}$    &  $d^7$ &  $3.2\times 10^6$  \\
     Ni$^{2+}$    &  $d^8$ &  $3.2\times 10^4$  \\
     V$^{3+}$     &  $d^2$ &  $5.0\times 10^2$  \\
     Fe$^{3+}$    &  $d^5$ &  $1.6\times 10^2$  \\
     Cr$^{3+}$    &  $d^3$ &  $2.4\times 10^{-6}$  \\\hline
    \end{tabular}
\end{table}

\begin{figure}
    \centering
    \includegraphics[width=0.9\textwidth]{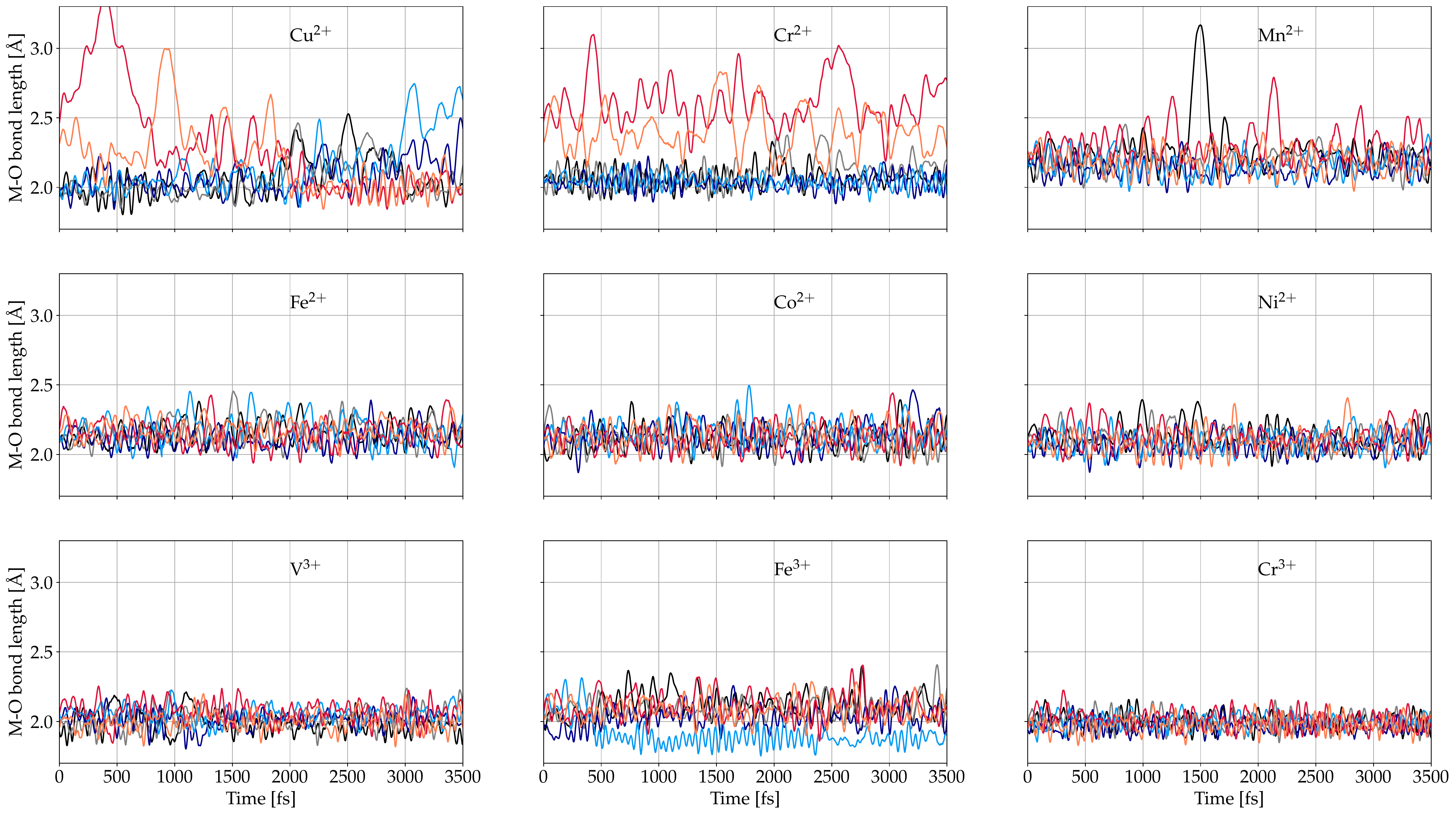}
    \caption{Metal-oxygen bond distance fluctuations along the full \emph{ab initio} molecular dynamics trajectories for all transition metal ions considered in this study. Each colored line represents one oxygen atom of the first solvation shell.}
    \label{fig:bondsall}
\end{figure}

In order to evaluate the effect that the labile nature of this complexes have on the vertical IPs,
we ran \emph{ab initio} molecular 
dynamics (AIMD) simulations of the implicitly solvated \ce{[M(H2O)40]^n+} models at 298.15 K.  All collected trajectories were, at least, 3.5 ps long. The size of the metal-oxygen bond distances along the trajectory, shown in Figure \ref{fig:bondsall}, correlates rather well with the experimental water exchange rates. In particular, both \ce{[Cu(H2O)40]^{2+}} and \ce{[Cr(H2O)40]^{2+}} exhibit oscillations larger than 0.5 \AA over the whole trajectory, but \ce{[Cr(H2O)40]^{2+}} only shows jumps in the distortion axis similar to Berry pseudorotations (see below for more details on the \ce{[Cr(H2O)40]^{2+}} dynamics). The $\ce{[Mn(H2O)40]^{2+}}$ complex also undergoes significant fluctuations in the metal-oxygen distance, but these are less frequent and smaller overall compared with the hydrated \ce{[Cu]^{2+}} and \ce{[Cr]^{2+}} ions.

%copper(II) and chromium(II) ions.

\begin{figure}
    \centering
    \includegraphics[width=0.9\textwidth]{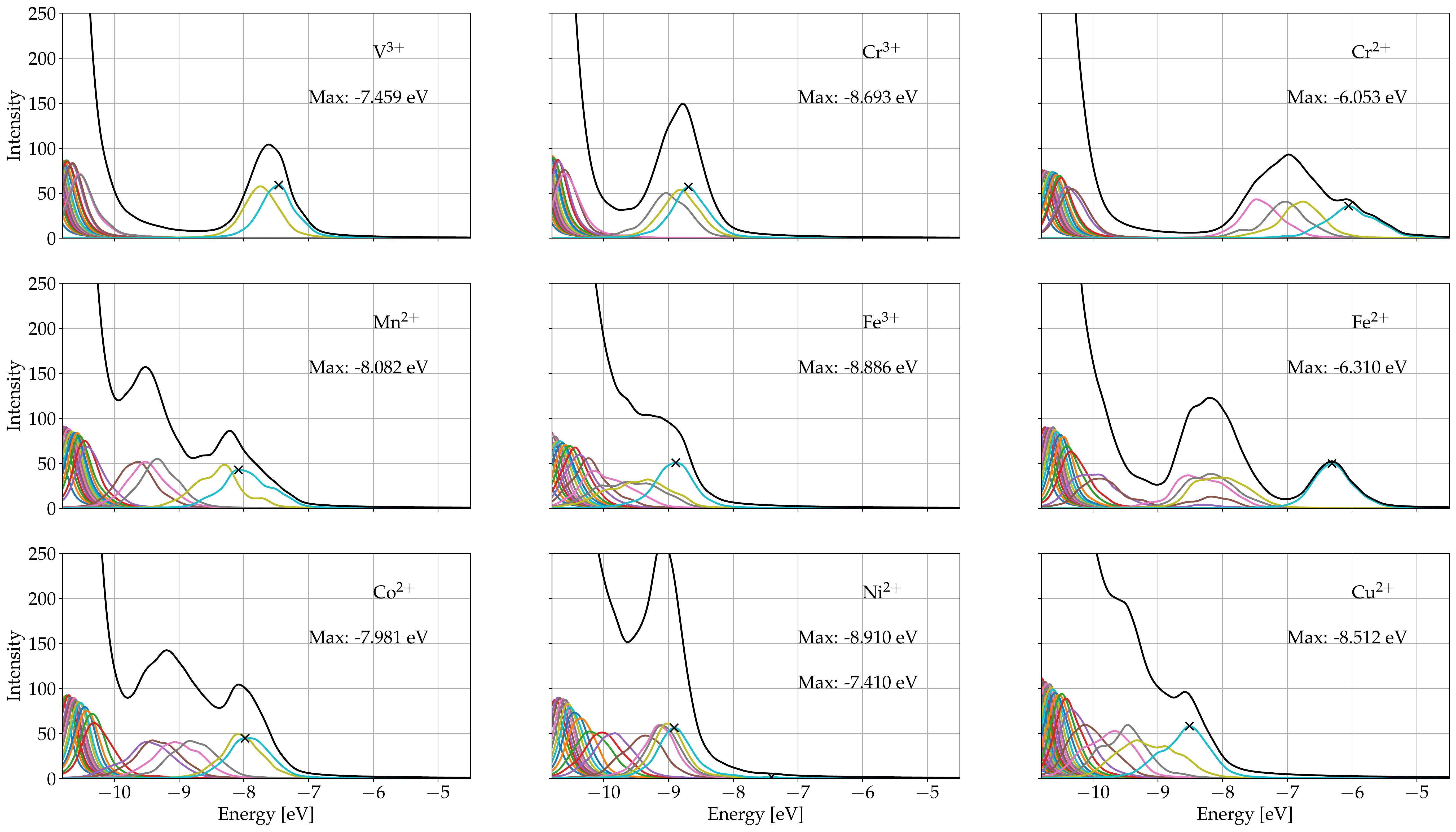}
    \caption{Simulated valence photoelectron spectra obtained by broadening the $G_0W_0$@r$^2$SCAN removal energies closest to the Fermi level. A total of 2400 removal energies per ion are included in the plots (120 snapshots, 20 energies per snapshot). The crosses mark local maxima. }
    \label{fig:pesall}
\end{figure}

The simulated valence photoemission spectra (PES) shown in Figure \ref{fig:pesall} were obtained by sampling 120 configurations from the last 2.5 ps of each trajectory. Twenty $G_0W_0$@r$^2$SCAN removal energies were obtained for each configuration. Each one of the 2400 removal energies was broadened with a pseudo-Voigt function\cite{Wertheim:1974:1369} with 0.25 eV full width at half maximum (FWHM). In Figure \ref{fig:pesall}, the contributions stemming from each individual hole state are depicted in lines with different colors, while the total spectrum is given in black. The local maxima of the low energy peaks are marked with crosses to facilitate comparison to the static results. As expected from the large metal-oxygen distance fluctuations, the HOMO peaks from \ce{Cu^{2+}}, \ce{Cr^{2+}}, and \ce{Mn^{2+}} show wider dispersion ( $>$ 2.0 eV) than all other ions ( $<$ 1.5 eV). However, all HOMO peaks spread for 1 eV at least, even for ions with narrow metal-oxygen distances oscillations (as is the case for all 3+ complexes). Table \ref{tab:comp} compares the triple-$\zeta$ $G_0W_0$@r$^2$SCAN IPs obtained using the optimized geometries for each complex with those from the local maxima of the PES in Figure \ref{fig:pesall}. The IPs extracted from the simulated PES are, in general, larger for the divalent cations and smaller for the trivalent ones, the sole exception being \ce{Cu^{2+}}. Overall, a 0.4 eV shift was observed, but three big outliers contribute most to this average: \ce{Fe^{3+}}, \ce{Fe^{2+}}, and \ce{Cr^{2+}}. 

\begin{table}[]
    \centering
    \caption{Comparison between the $G_0W_0$@r$^2$SCAN IPs [eV] obtained using the optimized geometries for each ion and the local maxima of the simulated photoemission spectra. All calculations were performed using the triple-$\zeta$ combination described in the main text.}
    \label{tab:comp}
    \begin{tabular}{c c c c}
                             & Exptl.  & Optimized & Local Maxima \\\hline
       \ce{[V(H2O)40]^{3+}}  & 8.41    &  7.57      & 7.46 \\
       \ce{[Cr(H2O)40]^{3+}} & 9.48    &  8.81      & 8.69 \\
       \ce{[Cr(H2O)40]^{2+}} & 6.76    &  5.36      & 6.05 \\
       \ce{[Mn(H2O)40]^{2+}} & 8.82    &  7.83      & 8.08 \\
       \ce{[Fe(H2O)40]^{3+}} & 10.28    & 10.10     & 8.89 \\
       \ce{[Fe(H2O)40]^{2+}} & 7.13    &  5.74     & 6.31 \\
       \ce{[Co(H2O)40]^{2+}} & 8.70    &  7.63     & 7.98 \\
       \ce{[Ni(H2O)40]^{2+}} & 9.45    &  8.76     & 8.91 \\
       \ce{[Cu(H2O)40]^{2+}} & 9.65    &  8.74     & 8.51 \\\hline
    \end{tabular}
\end{table}

\subsection{A deeper look into the \ce{[Cu(H2O)40]^{2+}} dynamics}
\begin{figure}
 \includegraphics[width=0.95\textwidth]{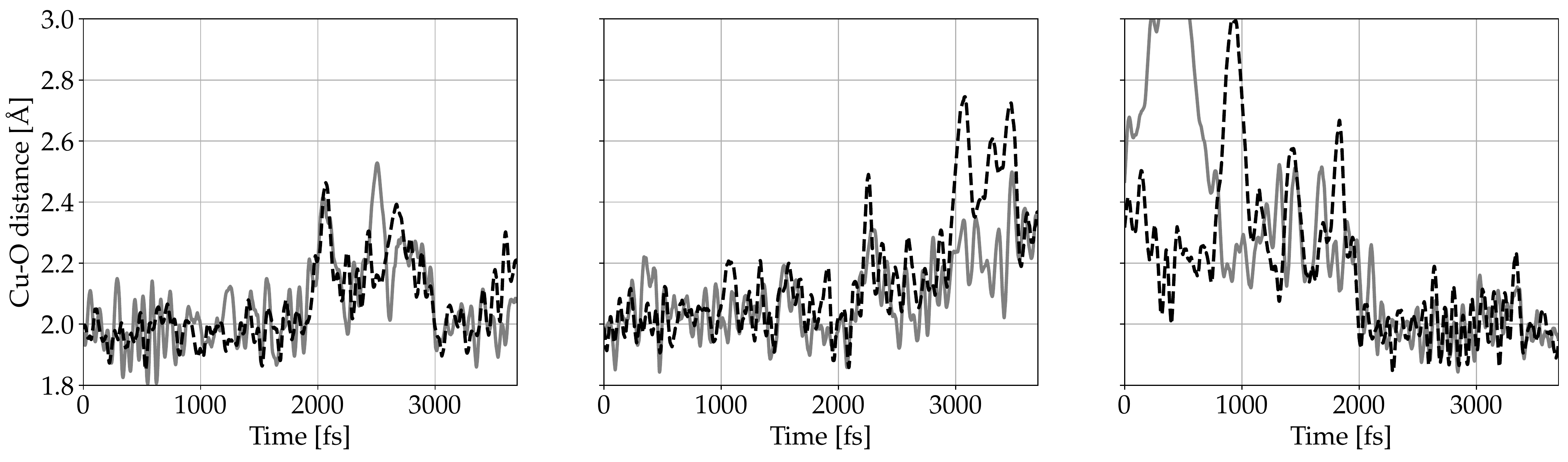}
 \includegraphics[width=0.95\textwidth]{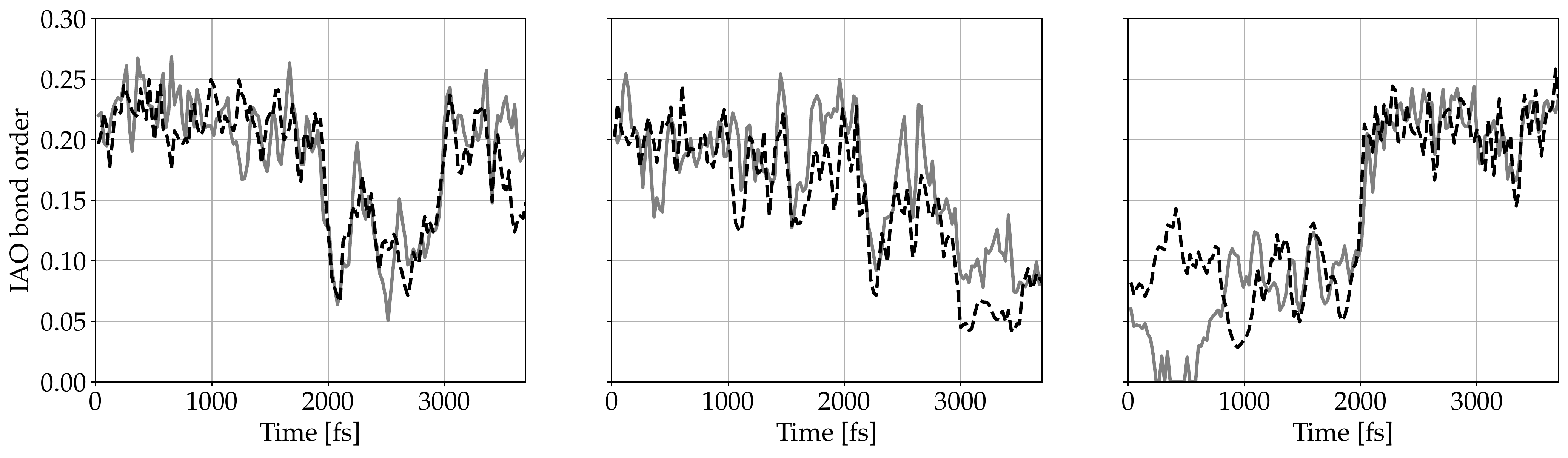}
 \caption{Cu--O bonds distances [\AA] (top) and bond orders (bottom) for the six nearest oxygen atoms. The plots were split by axial pairs in order to facilitate visualization, with each column corresponding to a given pair. Note that the ordinate of the top panel was cut at 3 \AA\ in order to reduce white space. The bond orders were obtained using intrinsic atomic orbitals (IAOs) representation.}
 \label{fig:cu2pdist}
\end{figure}

 It has been customary to define the different coordination symmetries probably present in the hydrated \ce{Cu^{2+}} ion as 6, 5+1, 4+1+1, and 5-coordinated complexes. Sharp distinctions between
the different symmetries were not observed, but rather
complex dynamics were obtained. 
Figure \ref{fig:cu2pdist} shows the Cu--O bond distance fluctuations
for the six nearest oxygen atoms. Equatorial
bond lengths fluctuate around a mean value of slightly less than 2 \AA. Axial
bond lengths exhibit wider oscillations around 2.2--2.5 \AA. 
In some instances, however, axial molecules can be found at distances greater than 2.75 \AA. 
Note also the presence of Berry-like pseudorotations
every $\sim 1$ ps: each pair of water molecules (corresponding to each column in Figure \ref{fig:cu2pdist})
occupied the axial position at different times, with the first pair (left column) occupying the position
between 2 and 3 ps, the second pair (middle column) from the 3 ps mark onwards, and the last pair (right column) during the first 2 ps. These pseudorotations were previously found in
MD simulations, using periodic boundary conditions, 
at about the same frequency \cite{Pasquarello:2001:856}.

The axial waters were further classified as bonded, partially bonded (corresponding to the ``+1'' in the 5+1, and 4+1+1 coordination symmetries), or not bonded by using Wiberg bond orders derived from intrinsic atomic orbitals (IAOs)\cite{Knizia:2013}. The bond order fluctuations shown in Figure \ref{fig:cu2pdist} are relatively larger than the corresponding fluctuations in the Cu--O distance. This allowed the discrimination of the different coordination symmetries. The region around IAO bond order of 0.20 corresponds to equatorial water molecules, while the region around 0.09 corresponds to axial ones. Based on this, we define water molecules with bond orders $> 0.09$ as fully bonded, those with bond orders between $0.05$ and $0.09$ as partially bonded, and those with bond orders $< 0.05$ as not bonded. Using this classification, we observe the following distribution of coordination modes for \ce{[Cu(H2O)40]^2+}: 1) a $4+1$ coordination (2\%); 2) $4+1+1$ coordination (18.3\%); 3) $5$ coordination (16.3\%); $5+1$ coordination (37.3\%); and 4) $6$ coordination (26\%). For comparison, recent experimental results obtained by combining EXAFS and MXAN found that Cu$^{2+}_{(\text{aq})}$ existed half of the time as an axially elongated square pyramidal (5-coordinated) structure and the other half as a $5+1$ structure\cite{Frank:2018:204302}. 

\begin{figure}
    \includegraphics[width=0.75\textwidth]{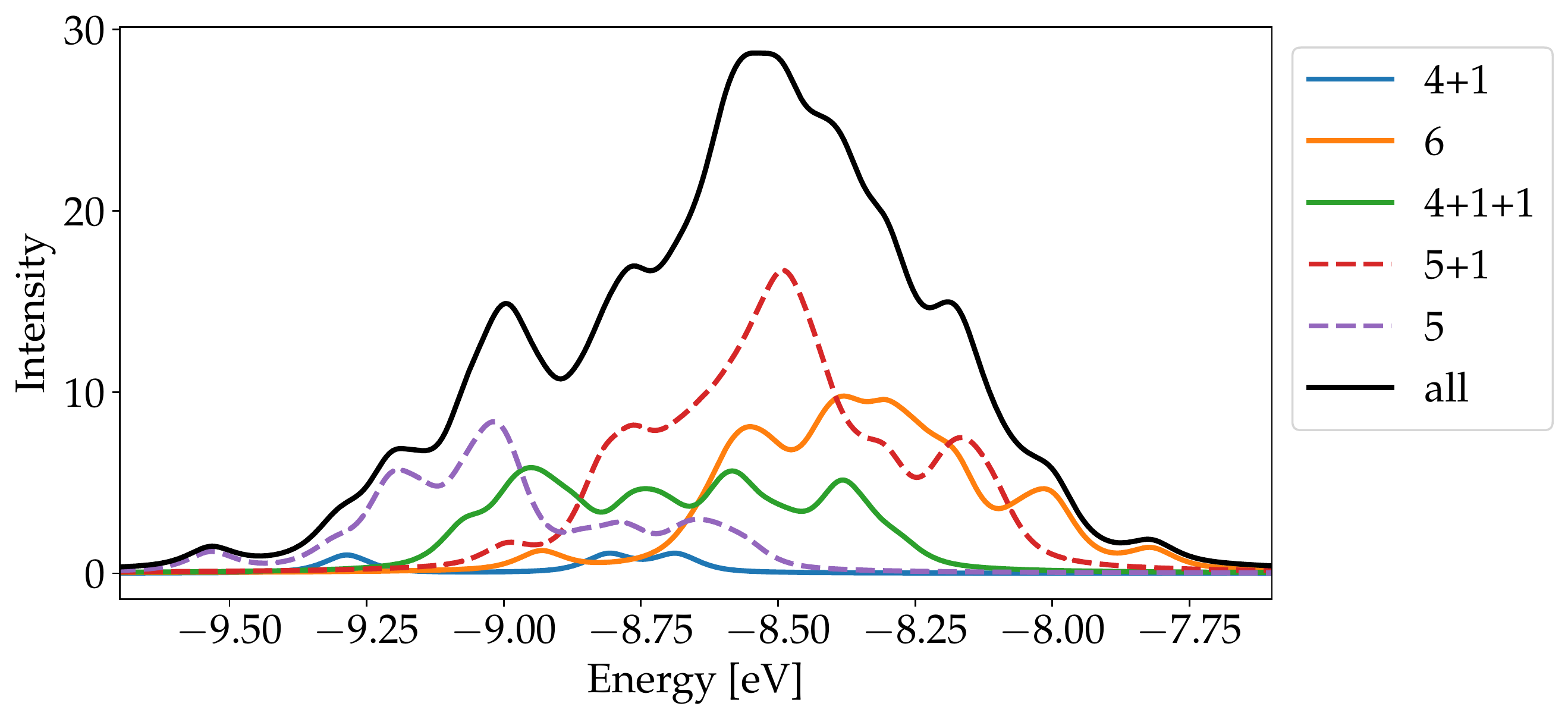} \\
    \includegraphics[width=0.75\textwidth]{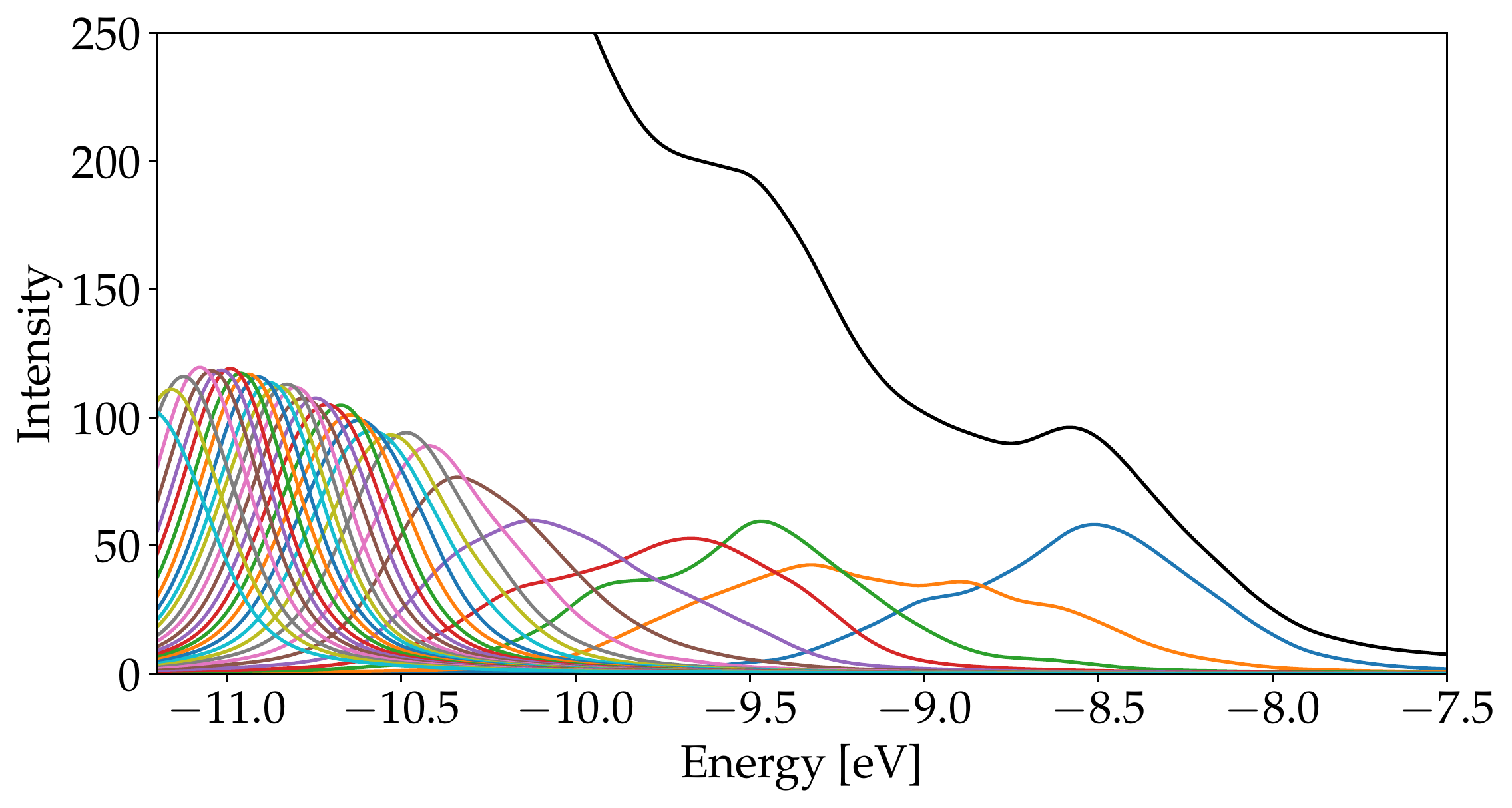} \\
    \includegraphics[width=0.75\textwidth]{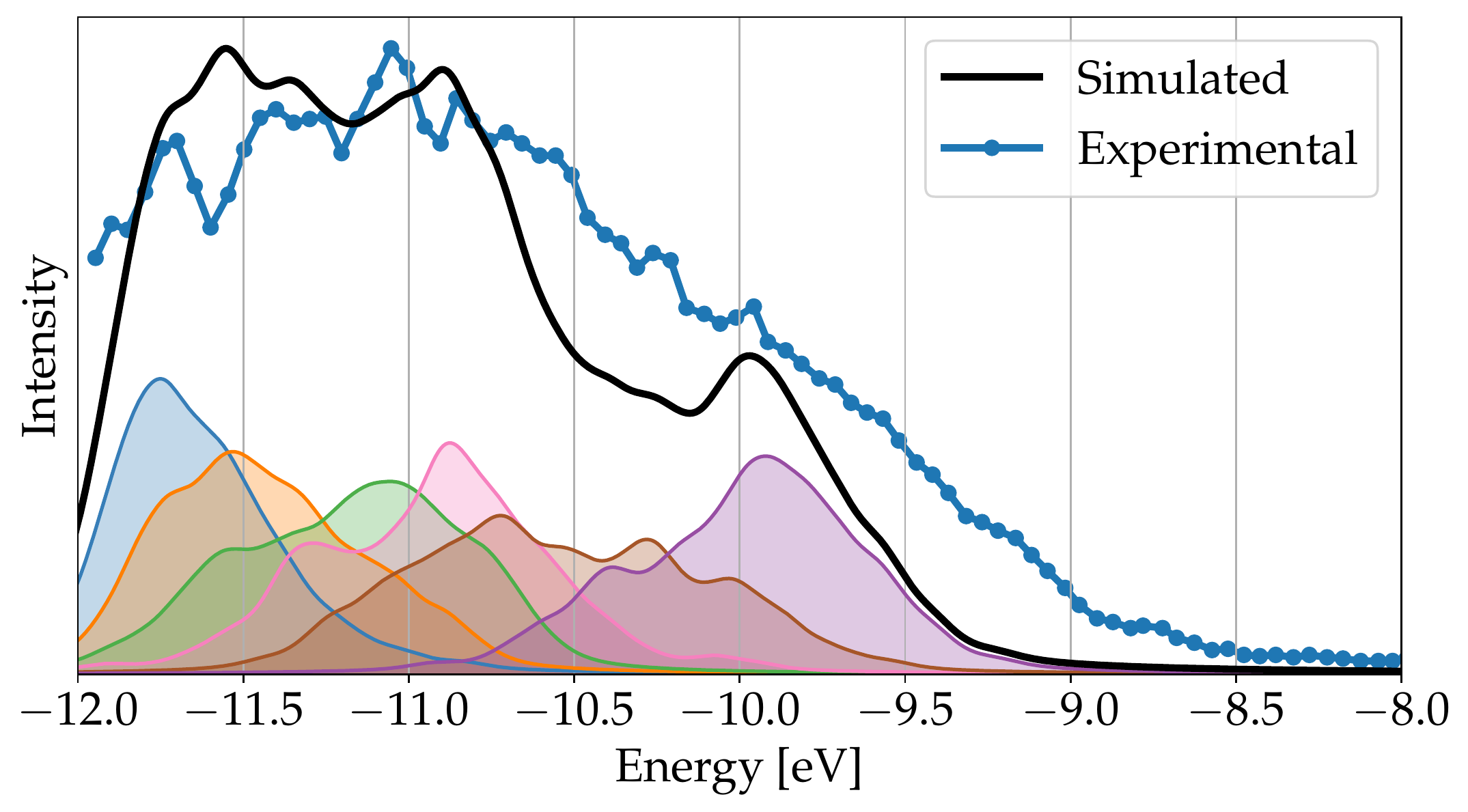}
    \caption{Top: changes in the highest occupied quasiparticle energy [eV] along a 3.7 ps AIMD trajectory of the hydrated \ce{Cu^2+} ion. The plot was obtained by broadening each peak with a Lorentzian with 0.1 eV FWHM. Middle: Total valence photoelectron spectra of hydrated \ce{Cu^2+} along the same 3.7 ps trajectory. Individual contributions from the 30 highest occupied quasiparticle states were broadened with a 0.25 eV FWHM Lorentzian. {\color{blue} Bottom: Comparison of the simulated and experimental valence photoelectron spectrum (from Ref. \citenum{Yepes:2014:6850}) of the hydrated \ce{Cu^2+} ion. The simulated spectrum was shifted by 1.4 eV and excluded ionizations from water orbitals.}}
    \label{fig:homo}
\end{figure}

Four dominant conformations Cu$^{2+}_{(\text{aq})}$, according to our classification (4+1+1, 5, 5+1, 6), have a significant impact on the highest occupied quasiparticle (HOQP) energy of \ce{[Cu(H2O)40]^2+}. The top panel of Figure \ref{fig:homo} shows, in black, the contribution of the HOQP to the valence photoelectron spectrum of Cu$^{2+}_{(\text{aq})}$. Note that the range of HOQP energies sampled is about 2.0 eV, which is rather large. The top panel of Figure \ref{fig:homo} also shows the contribution stemming from each individual coordination mode. As expected, lower coordination modes, according to our classification, lead to larger vertical IPs. The total spectra in the top panel of Figure \ref{fig:homo} suggest
sensitivity to the existence of at least three major structures (5, 5+1, and 6) in Cu$^{2+}_{(\text{aq})}$.

The {\color{blue} middle} panel of Figure \ref{fig:homo} shows the total simulated valence PES in a 4 eV window (30 removal energies). The spectrum includes contributions from water, which bundle together from -11.5 eV to -10.5 eV. The rest of the spectrum, arising from the $d$ shell of \ce{Cu^2+}, is included within a 3 eV span between -10.5 eV and -7.5 eV. This is roughly the same width shown in the differential valence PES of ref. \citenum{Yepes:2014:6850}, {\color{blue} whose digitized version is shown in the bottom panel of Figure \ref{fig:homo}. The simulated (shifted by 1.4 eV) and experimental spectra have} roughly the same characteristics, including a shoulder {\color{blue} around -10.1 eV} and a maximum around {\color{blue} -11.0 eV}. 

This finding reinforces the suggestion that more than two solvation shells are needed in order to simulate the complex dynamics of the \ce{Cu^2+} ion in an aqueous solution. A short constant temperature AIMD simulation of \ce{[Cu(H2O)18]^n+} also corroborates our suggestion, as the structural dynamics were not as complex as the one for \ce{[Cu(H2O)40]^n+}. {\color{blue} Given the timescale of the pseudorotations, it is likely that a longer AIMD trajectory would end up broadening the simulated spectra, improving the agreement to the experimental one}.

\section{Conclusion}

In summary, we have presented a detailed computational study of the
$G_0W_0$@DFT vertical IPs of the first-row TM aqua ions with comparisons to
experimental data.  Vertical IPs were calculated on a
range of TM ion clusters of increasing size 
(using 6, 18, 40, and 60 explicit water molecules) surrounded by a continuum solvation model. We determined that more than 2 solvation shells (18 water molecules) are needed to obtain converged results, but there seems to be no need to go beyond 40 explicit water molecules. The convergence with respect to the basis set size was also assessed using a complete basis set extrapolation scheme based on the Riemann $\zeta$ function. The quadruple-$\zeta$ IPs still underestimate the basis set limit, obtained with the $\zeta(3)$ approach, by about 0.5 eV. The extrapolated vertical $G_0W_0$ IPs show reasonably good agreement to the experimental values (0.13 eV underestimation) when using r$^2$SCAN orbitals as a starting point. 
 We also evaluated the
performance of two other starting points, namely, PBE, and PBE0. The PBE starting point leads to a severe underestimation of the IP values, while the PBE0 starting point will likely lead to overestimation in the basis set limit.

We have assessed the influence that different conformations have on the $G_0W_0$ IP values by performing molecular dynamics simulations of all clusters with 40 explicit water molecules. The vertical IPs varied more than 1 eV as a function of the geometrical configuration. In some cases (\ce{Cu^{2+}}, \ce{Cr^{2+}}, and \ce{Mn^{2+}}), the IPs varied more than 2 eV. These variations correlate well with experimental water exchange rates. {\color{blue} Moreover, we have shown that a simulated $G_0W_0$ valence photoelectron spectrum, taken from the snapshots of the molecular dynamics simulations, shows the same characteristics as the experimental spectrum, including width and the position of shoulders and local maxima. This suggests that DFT can correctly describe the complex dynamics of these aqua ions, and that $G_0W_0$ is accurate enough to reproduce the experimental observations.}

Utilizing water as the common solvation ligand for all the
TM ions makes this series an excellent benchmark to test the
predictive ability of theory not only from a structural standpoint, but also for spectroscopies and for a deeper understanding of complex ligand chemistries. We have demonstrated that the $G_0W_0$ approximation offers a good balance between computational cost and accuracy for predicting the vertical IPs of hydrated open-shell first-row transition metal cations and for capturing the effects of the complex structural dynamics that some of the cations exhibit in aqueous solution.

Finally, as an outlook, we note that other (partial) self-consistent methods of the $GW$ hierarchy may offer improvements, but at an increased computational cost. The $G_0W_0$ results presented here can be used to test the quality of those methods.

%omplexes and reproducing the complex dynamics that some of %them show in aqueous solution.

\suppinfo
Details about modifications to the COSMO solvation model implementation, detailed optimized metal-oxygen bond distances, and individual vertical IPs can be found in the Supporting Information file.

\begin{acknowledgement}

D.M.R, E.A., N.G. acknowledge support from the Center for Scalable Predictive methods for Excitations and Correlated phenomena (SPEC), which is funded as part of the Computational Chemical Sciences (CCS) program under FWP 70942 and J.L.F is supported under FWP 16248. Both projects are funded by the U.S. Department of Energy, Office of Science, Office of Basic Energy Sciences, Division of Chemical Sciences, Geosciences and Biosciences at Pacific Northwest National Laboratory (PNNL). PNNL is a multi-program national laboratory operated by Battelle Memorial Institute for the United States Department of Energy under DOE contract number DE-AC05-76RL01830. This research benefited from computational resources provided by EMSL, a DOE Office of Science User Facility sponsored by the Office of Biological and Environmental Research and located at PNNL. This research also used resources of the National Energy Research Scientific Computing Center (NERSC), a U.S. Department of Energy Office of Science User Facility located at Lawrence Berkeley National Laboratory, operated under Contract No. DE-AC02-05CH11231.
\end{acknowledgement}

\bibliography{bibliography.bib,gw.bib}

\providecommand{\latin}[1]{#1}
\makeatletter
\providecommand{\doi}
  {\begingroup\let\do\@makeother\dospecials
  \catcode`\{=1 \catcode`\}=2 \doi@aux}
\providecommand{\doi@aux}[1]{\endgroup\texttt{#1}}
\makeatother
\providecommand*\mcitethebibliography{\thebibliography}
\csname @ifundefined\endcsname{endmcitethebibliography}
  {\let\endmcitethebibliography\endthebibliography}{}
\begin{mcitethebibliography}{70}
\providecommand*\natexlab[1]{#1}
\providecommand*\mciteSetBstSublistMode[1]{}
\providecommand*\mciteSetBstMaxWidthForm[2]{}
\providecommand*\mciteBstWouldAddEndPuncttrue
  {\def\EndOfBibitem{\unskip.}}
\providecommand*\mciteBstWouldAddEndPunctfalse
  {\let\EndOfBibitem\relax}
\providecommand*\mciteSetBstMidEndSepPunct[3]{}
\providecommand*\mciteSetBstSublistLabelBeginEnd[3]{}
\providecommand*\EndOfBibitem{}
\mciteSetBstSublistMode{f}
\mciteSetBstMaxWidthForm{subitem}{(\alph{mcitesubitemcount})}
\mciteSetBstSublistLabelBeginEnd
  {\mcitemaxwidthsubitemform\space}
  {\relax}
  {\relax}

\bibitem[Crans and Kostenkova(2020)Crans, and Kostenkova]{Crans:2020:104}
Crans,~D.~C.; Kostenkova,~K. Open questions on the biological roles of
  first-row transition metals. \emph{Communications Chemistry} \textbf{2020},
  \emph{3}, 104\relax
\mciteBstWouldAddEndPuncttrue
\mciteSetBstMidEndSepPunct{\mcitedefaultmidpunct}
{\mcitedefaultendpunct}{\mcitedefaultseppunct}\relax
\EndOfBibitem
\bibitem[Stumm and Sulzberger(1992)Stumm, and Sulzberger]{Stumm_1992}
Stumm,~W.; Sulzberger,~B. The Cycling of Iron in Natural Environments:
  Considerations based on Laboratory Studies of Heterogeneous Redox Processes.
  \emph{Geochim. Cosmochim. Acta} \textbf{1992}, \emph{56}, 3233--3257\relax
\mciteBstWouldAddEndPuncttrue
\mciteSetBstMidEndSepPunct{\mcitedefaultmidpunct}
{\mcitedefaultendpunct}{\mcitedefaultseppunct}\relax
\EndOfBibitem
\bibitem[Richens(1997)]{Richens:1997}
Richens,~D.~T. \emph{The Chemistry of Aqua Ions: Synthesis, Structure, and
  Reactivity: A Tour through the Periodic Table of the Elements}; Wiley New
  York, 1997\relax
\mciteBstWouldAddEndPuncttrue
\mciteSetBstMidEndSepPunct{\mcitedefaultmidpunct}
{\mcitedefaultendpunct}{\mcitedefaultseppunct}\relax
\EndOfBibitem
\bibitem[Bhattacharjee \latin{et~al.}(2022)Bhattacharjee, Isegawa,
  Garcia-Rat{\'e}s, Neese, and Pantazis]{Bhattacharjee:2022:1619}
Bhattacharjee,~S.; Isegawa,~M.; Garcia-Rat{\'e}s,~M.; Neese,~F.;
  Pantazis,~D.~A. Ionization energies and redox potentials of hydrated
  transition metal ions: evaluation of domain-based local pair natural orbital
  coupled cluster approaches. \emph{J. Chem. Theory Comput.} \textbf{2022},
  \emph{18}, 1619--1632\relax
\mciteBstWouldAddEndPuncttrue
\mciteSetBstMidEndSepPunct{\mcitedefaultmidpunct}
{\mcitedefaultendpunct}{\mcitedefaultseppunct}\relax
\EndOfBibitem
\bibitem[Yepes \latin{et~al.}(2014)Yepes, Seidel, Winter, Blumberger, and
  Jaque]{Yepes:2014:6850}
Yepes,~D.; Seidel,~R.; Winter,~B.; Blumberger,~J.; Jaque,~P. Photoemission
  spectra and density functional theory calculations of 3d transition
  metal-aqua complexes (Ti-Cu) in aqueous solution. \emph{J. Phys. Chem. B}
  \textbf{2014}, \emph{118}, 6850--6863\relax
\mciteBstWouldAddEndPuncttrue
\mciteSetBstMidEndSepPunct{\mcitedefaultmidpunct}
{\mcitedefaultendpunct}{\mcitedefaultseppunct}\relax
\EndOfBibitem
\bibitem[Klamt and Sch{\"u}{\"u}rmann(1993)Klamt, and
  Sch{\"u}{\"u}rmann]{klamt1993}
Klamt,~A.; Sch{\"u}{\"u}rmann,~G. COSMO: A New Approach to Dielectric Screening
  in Solvents with Explicit Expressions for the Screening Energy and its
  Gradient. \emph{J. Chem. Soc., Perkin Trans. 2} \textbf{1993}, 799--805\relax
\mciteBstWouldAddEndPuncttrue
\mciteSetBstMidEndSepPunct{\mcitedefaultmidpunct}
{\mcitedefaultendpunct}{\mcitedefaultseppunct}\relax
\EndOfBibitem
\bibitem[Ghosh \latin{et~al.}(2022)Ghosh, Agarwal, Galib, Tran,
  Balasubramanian, Singh, Fulton, and Govind]{TM-XANES}
Ghosh,~S.; Agarwal,~H.; Galib,~M.; Tran,~B.; Balasubramanian,~M.; Singh,~N.;
  Fulton,~J.~L.; Govind,~N. Near-Quantitative Predictions of the First-Shell
  Coordination Structure of Hydrated First-Row Transition Metal Ions Using
  K-Edge X-ray Absorption Near-Edge Spectroscopy. \emph{The Journal of Physical
  Chemistry Letters} \textbf{2022}, \emph{13}, 6323--6330\relax
\mciteBstWouldAddEndPuncttrue
\mciteSetBstMidEndSepPunct{\mcitedefaultmidpunct}
{\mcitedefaultendpunct}{\mcitedefaultseppunct}\relax
\EndOfBibitem
\bibitem[Ghosh \latin{et~al.}()Ghosh, Mukamel, and Govind]{Ghosh:2023:5203}
Ghosh,~S.; Mukamel,~S.; Govind,~N. A Combined Wave Function and Density
  Functional Approach for K-Edge X-ray Absorption Near-Edge Spectroscopy: A
  Case Study of Hydrated First-Row Transition Metal Ions. \emph{The Journal of
  Physical Chemistry Letters} \emph{14}, 5203--5209\relax
\mciteBstWouldAddEndPuncttrue
\mciteSetBstMidEndSepPunct{\mcitedefaultmidpunct}
{\mcitedefaultendpunct}{\mcitedefaultseppunct}\relax
\EndOfBibitem
\bibitem[Hedin(1965)]{Hedin:1965:A796}
Hedin,~L. New method for calculating the one-particle Green's function with
  application to the electron-gas problem. \emph{Phys. Rev.} \textbf{1965},
  \emph{139}, A796--A823\relax
\mciteBstWouldAddEndPuncttrue
\mciteSetBstMidEndSepPunct{\mcitedefaultmidpunct}
{\mcitedefaultendpunct}{\mcitedefaultseppunct}\relax
\EndOfBibitem
\bibitem[Mejia-Rodriguez \latin{et~al.}(2021)Mejia-Rodriguez, Kuntisa, Apr\`a,
  and Govind]{Mejia-Rodriguez:2021:7504}
Mejia-Rodriguez,~D.; Kuntisa,~A.; Apr\`a,~E.; Govind,~N. Scalable molecular
  $GW$ calculations: Valence and core spectra. \emph{J. Chem. Theory Comput.}
  \textbf{2021}, \emph{17}, 7504--7517\relax
\mciteBstWouldAddEndPuncttrue
\mciteSetBstMidEndSepPunct{\mcitedefaultmidpunct}
{\mcitedefaultendpunct}{\mcitedefaultseppunct}\relax
\EndOfBibitem
\bibitem[Mejia-Rodriguez \latin{et~al.}(2022)Mejia-Rodriguez, Kunitsa,
  Apr{\`a}, and Govind]{Mejia-Rodriguez:2022:4919}
Mejia-Rodriguez,~D.; Kunitsa,~A.; Apr{\`a},~E.; Govind,~N. Basis Set Selection
  for Molecular Core-Level GW Calculations. \emph{J. Chem. Theory Comput.}
  \textbf{2022}, \emph{18}, 4919--4926\relax
\mciteBstWouldAddEndPuncttrue
\mciteSetBstMidEndSepPunct{\mcitedefaultmidpunct}
{\mcitedefaultendpunct}{\mcitedefaultseppunct}\relax
\EndOfBibitem
\bibitem[Aryasetiawan and Gunnarsson(1998)Aryasetiawan, and
  Gunnarsson]{Aryasetiawan:1998:237}
Aryasetiawan,~F.; Gunnarsson,~O. The GW method. \emph{Rep. Prog. Phys.}
  \textbf{1998}, \emph{61}, 237--312\relax
\mciteBstWouldAddEndPuncttrue
\mciteSetBstMidEndSepPunct{\mcitedefaultmidpunct}
{\mcitedefaultendpunct}{\mcitedefaultseppunct}\relax
\EndOfBibitem
\bibitem[Onida \latin{et~al.}(2002)Onida, Reining, and Rubio]{Onida:2002:601}
Onida,~G.; Reining,~L.; Rubio,~A. Electronic excitations: density-functional
  versus many-body Green’s-function approaches. \emph{Rev. Mod. Phys.}
  \textbf{2002}, \emph{74}, 601--629\relax
\mciteBstWouldAddEndPuncttrue
\mciteSetBstMidEndSepPunct{\mcitedefaultmidpunct}
{\mcitedefaultendpunct}{\mcitedefaultseppunct}\relax
\EndOfBibitem
\bibitem[Martin \latin{et~al.}(2016)Martin, Reining, and Ceperley]{Martin:2016}
Martin,~R.~M.; Reining,~L.; Ceperley,~D.~M. \emph{Interacting electrons};
  Cambridge University Press, 2016\relax
\mciteBstWouldAddEndPuncttrue
\mciteSetBstMidEndSepPunct{\mcitedefaultmidpunct}
{\mcitedefaultendpunct}{\mcitedefaultseppunct}\relax
\EndOfBibitem
\bibitem[Reining(2018)]{Reining:2018:e1344}
Reining,~L. The GW approximation: content, successes and limitations.
  \emph{Wiley Interdisciplinary Reviews: Computational Molecular Science}
  \textbf{2018}, \emph{8}, e1344\relax
\mciteBstWouldAddEndPuncttrue
\mciteSetBstMidEndSepPunct{\mcitedefaultmidpunct}
{\mcitedefaultendpunct}{\mcitedefaultseppunct}\relax
\EndOfBibitem
\bibitem[Rostgaard \latin{et~al.}(2010)Rostgaard, Jacobsen, and
  Thygesen]{Rostgaard:2010:085103}
Rostgaard,~C.; Jacobsen,~K.~W.; Thygesen,~K.~S. Fully self-consistent GW
  calculations for molecules. \emph{Phys. Rev. B} \textbf{2010}, \emph{81},
  085103\relax
\mciteBstWouldAddEndPuncttrue
\mciteSetBstMidEndSepPunct{\mcitedefaultmidpunct}
{\mcitedefaultendpunct}{\mcitedefaultseppunct}\relax
\EndOfBibitem
\bibitem[Blase \latin{et~al.}(2011)Blase, Attaccalite, and
  Olevano]{Blase:2011:115103}
Blase,~X.; Attaccalite,~C.; Olevano,~V. First-principles GW calculations for
  fullerenes, porphyrins, phtalocyanine, and other molecules of interest for
  organic photovoltaic applications. \emph{Phys. Rev. B} \textbf{2011},
  \emph{83}, 115103\relax
\mciteBstWouldAddEndPuncttrue
\mciteSetBstMidEndSepPunct{\mcitedefaultmidpunct}
{\mcitedefaultendpunct}{\mcitedefaultseppunct}\relax
\EndOfBibitem
\bibitem[Bruneval(2012)]{Bruneval2012}
Bruneval,~F. {Ionization energy of atoms obtained from GW self-energy or from
  random phase approximation total energies}. \emph{J. Chem. Phys.}
  \textbf{2012}, \emph{136}, 194107\relax
\mciteBstWouldAddEndPuncttrue
\mciteSetBstMidEndSepPunct{\mcitedefaultmidpunct}
{\mcitedefaultendpunct}{\mcitedefaultseppunct}\relax
\EndOfBibitem
\bibitem[Caruso \latin{et~al.}(2012)Caruso, Rinke, Ren, Scheffler, and
  Rubio]{Caruso:2012:081102}
Caruso,~F.; Rinke,~P.; Ren,~X.; Scheffler,~M.; Rubio,~A. Unified description of
  ground and excited states of finite systems: The self-consistent GW approach.
  \emph{Phys. Rev. B} \textbf{2012}, \emph{86}, 081102\relax
\mciteBstWouldAddEndPuncttrue
\mciteSetBstMidEndSepPunct{\mcitedefaultmidpunct}
{\mcitedefaultendpunct}{\mcitedefaultseppunct}\relax
\EndOfBibitem
\bibitem[van Setten \latin{et~al.}(2013)van Setten, Weigend, and
  Evers]{VanSetten2013}
van Setten,~M.~J.; Weigend,~F.; Evers,~F. {The $GW$-method for quantum
  chemistry applications: Theory and Implementation}. \emph{J. Chem. Theory
  Comput.} \textbf{2013}, \emph{9}, 232--246\relax
\mciteBstWouldAddEndPuncttrue
\mciteSetBstMidEndSepPunct{\mcitedefaultmidpunct}
{\mcitedefaultendpunct}{\mcitedefaultseppunct}\relax
\EndOfBibitem
\bibitem[Bruneval \latin{et~al.}(2016)Bruneval, Rangel, Hamed, Shao, Yang, and
  Neaton]{Bruneval2016}
Bruneval,~F.; Rangel,~T.; Hamed,~S.~M.; Shao,~M.; Yang,~C.; Neaton,~J.~B. molgw
  1: Many-body perturbation theory software for atoms, molecules, and clusters.
  \emph{Comput. Phys. Comm.} \textbf{2016}, \emph{208}, 149--161\relax
\mciteBstWouldAddEndPuncttrue
\mciteSetBstMidEndSepPunct{\mcitedefaultmidpunct}
{\mcitedefaultendpunct}{\mcitedefaultseppunct}\relax
\EndOfBibitem
\bibitem[van Setten \latin{et~al.}(2015)van Setten, Caruso, Sharifzadeh, Ren,
  Scheffler, Liu, Lischner, Lin, Deslippe, Louie, Yang, Weigend, Neaton, Evers,
  and Rinke]{vanSetten:2015:5665}
van Setten,~M.~J.; Caruso,~F.; Sharifzadeh,~S.; Ren,~X.; Scheffler,~M.;
  Liu,~F.; Lischner,~J.; Lin,~L.; Deslippe,~J.~R.; Louie,~S.~G. \latin{et~al.}
  $GW$100: Benchmarking $G_0W_0$ for molecular systems. \emph{J. Chem. Theory
  Comput.} \textbf{2015}, \emph{11}, 5665--5687\relax
\mciteBstWouldAddEndPuncttrue
\mciteSetBstMidEndSepPunct{\mcitedefaultmidpunct}
{\mcitedefaultendpunct}{\mcitedefaultseppunct}\relax
\EndOfBibitem
\bibitem[Golze \latin{et~al.}(2019)Golze, Dvorak, and Rinke]{Golze:2019:377}
Golze,~D.; Dvorak,~M.; Rinke,~P. The GW compendium: A practical guide to
  theoretical photoemission spectroscopy. \emph{Front. Chem.} \textbf{2019},
  \emph{7}, 377\relax
\mciteBstWouldAddEndPuncttrue
\mciteSetBstMidEndSepPunct{\mcitedefaultmidpunct}
{\mcitedefaultendpunct}{\mcitedefaultseppunct}\relax
\EndOfBibitem
\bibitem[Golze \latin{et~al.}(2020)Golze, Keller, and Rinke]{Golze:2020:1840}
Golze,~D.; Keller,~L.; Rinke,~P. Accurate absolute and relative core-level
  binding energies from $GW$. \emph{J. Phys. Chem. Letters} \textbf{2020},
  \emph{11}, 1840--1847\relax
\mciteBstWouldAddEndPuncttrue
\mciteSetBstMidEndSepPunct{\mcitedefaultmidpunct}
{\mcitedefaultendpunct}{\mcitedefaultseppunct}\relax
\EndOfBibitem
\bibitem[Adler(1962)]{Adler:1962}
Adler,~S.~L. Quantum theory of the dielectric constant in real solids.
  \emph{Phys. Rev.} \textbf{1962}, \emph{126}, 413--420\relax
\mciteBstWouldAddEndPuncttrue
\mciteSetBstMidEndSepPunct{\mcitedefaultmidpunct}
{\mcitedefaultendpunct}{\mcitedefaultseppunct}\relax
\EndOfBibitem
\bibitem[Wiser(1963)]{Wiser:1963}
Wiser,~N. Dielectric constant with local field effects included. \emph{Phys.
  Rev.} \textbf{1963}, \emph{129}, 62--69\relax
\mciteBstWouldAddEndPuncttrue
\mciteSetBstMidEndSepPunct{\mcitedefaultmidpunct}
{\mcitedefaultendpunct}{\mcitedefaultseppunct}\relax
\EndOfBibitem
\bibitem[Aprà \latin{et~al.}(2020)Aprà, Bylaska, de~Jong, Govind, Kowalski,
  Straatsma, Valiev, van Dam, Alexeev, Anchell, Anisimov, Aquino, Atta-Fynn,
  Autschbach, Bauman, Becca, Bernholdt, Bhaskaran-Nair, Bogatko, Borowski,
  Boschen, Brabec, Bruner, Cauët, Chen, Chuev, Cramer, Daily, Deegan, Dunning,
  Dupuis, Dyall, Fann, Fischer, Fonari, Früchtl, Gagliardi, Garza, Gawande,
  Ghosh, Glaesemann, Götz, Hammond, Helms, Hermes, Hirao, Hirata, Jacquelin,
  Jensen, Johnson, Jónsson, Kendall, Klemm, Kobayashi, Konkov, Krishnamoorthy,
  Krishnan, Lin, Lins, Littlefield, Logsdail, Lopata, Ma, Marenich, Martin~del
  Campo, Mejia-Rodriguez, Moore, Mullin, Nakajima, Nascimento, Nichols,
  Nichols, Nieplocha, Otero-de-la Roza, Palmer, Panyala, Pirojsirikul, Peng,
  Peverati, Pittner, Pollack, Richard, Sadayappan, Schatz, Shelton,
  Silverstein, Smith, Soares, Song, Swart, Taylor, Thomas, Tipparaju, Truhlar,
  Tsemekhman, Van~Voorhis, Vázquez-Mayagoitia, Verma, Villa, Vishnu,
  Vogiatzis, Wang, Weare, Williamson, Windus, Woliński, Wong, Wu, Yang, Yu,
  Zacharias, Zhang, Zhao, and Harrison]{NWChem}
Aprà,~E.; Bylaska,~E.~J.; de~Jong,~W.~A.; Govind,~N.; Kowalski,~K.;
  Straatsma,~T.~P.; Valiev,~M.; van Dam,~H. J.~J.; Alexeev,~Y.; Anchell,~J.
  \latin{et~al.}  NWChem: Past, present, and future. \emph{J. Chem. Phys.}
  \textbf{2020}, \emph{152}, 184102\relax
\mciteBstWouldAddEndPuncttrue
\mciteSetBstMidEndSepPunct{\mcitedefaultmidpunct}
{\mcitedefaultendpunct}{\mcitedefaultseppunct}\relax
\EndOfBibitem
\bibitem[Perdew \latin{et~al.}(1996)Perdew, Burke, and Ernzerhof]{pbe}
Perdew,~J.~P.; Burke,~K.; Ernzerhof,~M. Generalized gradient approximation made
  simple. \emph{Phys. Rev. Lett.} \textbf{1996}, \emph{77}, 3865--3868\relax
\mciteBstWouldAddEndPuncttrue
\mciteSetBstMidEndSepPunct{\mcitedefaultmidpunct}
{\mcitedefaultendpunct}{\mcitedefaultseppunct}\relax
\EndOfBibitem
\bibitem[Adamo and Barone(1999)Adamo, and Barone]{pbe0}
Adamo,~C.; Barone,~V. Toward Reliable Density Functional Methods Without
  Adjustable Parameters: The PBE0 Model. \emph{J. Chem. Phys.} \textbf{1999},
  \emph{110}, 6158--6170\relax
\mciteBstWouldAddEndPuncttrue
\mciteSetBstMidEndSepPunct{\mcitedefaultmidpunct}
{\mcitedefaultendpunct}{\mcitedefaultseppunct}\relax
\EndOfBibitem
\bibitem[Furness \latin{et~al.}(2020)Furness, Kaplan, Ning, Perdew, and
  Sun]{r2scan}
Furness,~J.~W.; Kaplan,~A.~D.; Ning,~J.; Perdew,~J.~P.; Sun,~J. Accurate and
  Numerically Efficient r2SCAN Meta-Generalized Gradient Approximation.
  \emph{J. Phys. Chem. Lett.} \textbf{2020}, \emph{11}, 8208--8215\relax
\mciteBstWouldAddEndPuncttrue
\mciteSetBstMidEndSepPunct{\mcitedefaultmidpunct}
{\mcitedefaultendpunct}{\mcitedefaultseppunct}\relax
\EndOfBibitem
\bibitem[Weigend \latin{et~al.}(2003)Weigend, Furche, and Ahlrichs]{def2}
Weigend,~F.; Furche,~F.; Ahlrichs,~R. Gaussian basis sets of quadruple zeta
  valence quality for atoms H-Kr. \emph{J. Chem. Phys.} \textbf{2003},
  \emph{119}, 12753--12762\relax
\mciteBstWouldAddEndPuncttrue
\mciteSetBstMidEndSepPunct{\mcitedefaultmidpunct}
{\mcitedefaultendpunct}{\mcitedefaultseppunct}\relax
\EndOfBibitem
\bibitem[Noro \latin{et~al.}(2012)Noro, Sekiya, and Koga]{sapporo}
Noro,~T.; Sekiya,~M.; Koga,~T. Segmented Contracted Basis Sets for Atoms H
  Through Xe: Sapporo-(DK)-nZP sets (n = D, T, Q). \emph{Theor. Chem. Acc.}
  \textbf{2012}, \emph{131}, 1124\relax
\mciteBstWouldAddEndPuncttrue
\mciteSetBstMidEndSepPunct{\mcitedefaultmidpunct}
{\mcitedefaultendpunct}{\mcitedefaultseppunct}\relax
\EndOfBibitem
\bibitem[Woon and Jr.(1995)Woon, and Jr.]{ccpcvnz1}
Woon,~D.~E.; Jr.,~T. H.~D. Gaussian basis sets for use in correlated molecular
  calculations. V. Core‐valence basis sets for boron through neon. \emph{J.
  Chem. Phys.} \textbf{1995}, \emph{103}, 4572--4585\relax
\mciteBstWouldAddEndPuncttrue
\mciteSetBstMidEndSepPunct{\mcitedefaultmidpunct}
{\mcitedefaultendpunct}{\mcitedefaultseppunct}\relax
\EndOfBibitem
\bibitem[Peterson and Jr.(2002)Peterson, and Jr.]{ccpcvnz2}
Peterson,~K.~A.; Jr.,~T. H.~D. Accurate correlation consistent basis sets for
  molecular core–valence correlation effects: The second row atoms Al–Ar,
  and the first row atoms B–Ne revisited. \emph{J. Chem. Phys.}
  \textbf{2002}, \emph{117}, 10548--10560\relax
\mciteBstWouldAddEndPuncttrue
\mciteSetBstMidEndSepPunct{\mcitedefaultmidpunct}
{\mcitedefaultendpunct}{\mcitedefaultseppunct}\relax
\EndOfBibitem
\bibitem[Bobalabanov and Peterson(2005)Bobalabanov, and Peterson]{ccpcvnz3}
Bobalabanov,~N.~B.; Peterson,~K.~A. Systematically convergent basis sets for
  transition metals. I. All-electron correlation consistent basis sets for the
  3d elements Sc–Zn. \emph{J. Chem. Phys.} \textbf{2005}, \emph{123},
  064107\relax
\mciteBstWouldAddEndPuncttrue
\mciteSetBstMidEndSepPunct{\mcitedefaultmidpunct}
{\mcitedefaultendpunct}{\mcitedefaultseppunct}\relax
\EndOfBibitem
\bibitem[Grimme \latin{et~al.}(2010)Grimme, Antony, Ehrlich, and Krieg]{DFTD3}
Grimme,~S.; Antony,~J.; Ehrlich,~S.; Krieg,~H. A Consistent and Accurate Ab
  Initio Parametrization of Density Functional Dispersion Correction (DFT-D)
  for the 94 Elements H-Pu. \emph{J. Chem. Phys.} \textbf{2010}, \emph{132},
  154104\relax
\mciteBstWouldAddEndPuncttrue
\mciteSetBstMidEndSepPunct{\mcitedefaultmidpunct}
{\mcitedefaultendpunct}{\mcitedefaultseppunct}\relax
\EndOfBibitem
\bibitem[Grimme \latin{et~al.}(2011)Grimme, Ehrlich, and Goerigk]{BJdamping}
Grimme,~S.; Ehrlich,~S.; Goerigk,~L. Effect of the damping function in
  dispersion corrected density functional theory. \emph{J. Comp. Chem.}
  \textbf{2011}, \emph{32}, 1456--1465\relax
\mciteBstWouldAddEndPuncttrue
\mciteSetBstMidEndSepPunct{\mcitedefaultmidpunct}
{\mcitedefaultendpunct}{\mcitedefaultseppunct}\relax
\EndOfBibitem
\bibitem[Ehlert \latin{et~al.}(2021)Ehlert, Huniar, Ning, Furness, Sun, Kaplan,
  Perdew, and Brandenburg]{r2scand4}
Ehlert,~S.; Huniar,~U.; Ning,~J.; Furness,~J.~W.; Sun,~J.; Kaplan,~A.~D.;
  Perdew,~J.~P.; Brandenburg,~J.~G. r$^2$SCAN-D4: Dispersion corrected
  meta-generalized gradient approximation for general chemical applications.
  \emph{J. Chem. Phys.} \textbf{2021}, \emph{154}, 061101\relax
\mciteBstWouldAddEndPuncttrue
\mciteSetBstMidEndSepPunct{\mcitedefaultmidpunct}
{\mcitedefaultendpunct}{\mcitedefaultseppunct}\relax
\EndOfBibitem
\bibitem[York and Karplus(1999)York, and Karplus]{york-karplus}
York,~D.~M.; Karplus,~M. A smooth solvation potential based on the
  conductor-like screening model. \emph{J. Phys. Chem. A} \textbf{1999},
  \emph{103}, 11060--11079\relax
\mciteBstWouldAddEndPuncttrue
\mciteSetBstMidEndSepPunct{\mcitedefaultmidpunct}
{\mcitedefaultendpunct}{\mcitedefaultseppunct}\relax
\EndOfBibitem
\bibitem[Dixon(1989)]{Dixon1989}
Dixon,~R. Spiral phyllotaxis. \emph{Computers Math. Applic.} \textbf{1989},
  \emph{17}, 535--538\relax
\mciteBstWouldAddEndPuncttrue
\mciteSetBstMidEndSepPunct{\mcitedefaultmidpunct}
{\mcitedefaultendpunct}{\mcitedefaultseppunct}\relax
\EndOfBibitem
\bibitem[Svergun(1994)]{Svergun1994}
Svergun,~D.~I. Solution scattering from biopolymers: advanced
  contrast-variation data analysis. \emph{Acta Cryst.} \textbf{1994},
  \emph{A50}, 391--402\relax
\mciteBstWouldAddEndPuncttrue
\mciteSetBstMidEndSepPunct{\mcitedefaultmidpunct}
{\mcitedefaultendpunct}{\mcitedefaultseppunct}\relax
\EndOfBibitem
\bibitem[Swinbank and Purser(2006)Swinbank, and Purser]{swinbank2006}
Swinbank,~R.; Purser,~R.~J. Fibonnaci grids: a novel approach to global
  modelling. \emph{Q. J. R. Meteorol. Soc.} \textbf{2006}, \emph{132},
  1769--1793\relax
\mciteBstWouldAddEndPuncttrue
\mciteSetBstMidEndSepPunct{\mcitedefaultmidpunct}
{\mcitedefaultendpunct}{\mcitedefaultseppunct}\relax
\EndOfBibitem
\bibitem[Gonz\'alez(2010)]{Gonzalez2010}
Gonz\'alez,~A. Measurement of areas on a sphere using Fibonacci and
  latitude-longitude lattices. \emph{Math. Geosci.} \textbf{2010}, \emph{42},
  49--64\relax
\mciteBstWouldAddEndPuncttrue
\mciteSetBstMidEndSepPunct{\mcitedefaultmidpunct}
{\mcitedefaultendpunct}{\mcitedefaultseppunct}\relax
\EndOfBibitem
\bibitem[Uudsemaa and Tamm(2003)Uudsemaa, and Tamm]{Uudsemaa2003}
Uudsemaa,~M.; Tamm,~T. Density-Functional Theory Calculations of Aqueous Redox
  Potentials of Fourth-Period Transition Metals. \emph{J. Phys. Chem. A}
  \textbf{2003}, \emph{107}, 9997--10003\relax
\mciteBstWouldAddEndPuncttrue
\mciteSetBstMidEndSepPunct{\mcitedefaultmidpunct}
{\mcitedefaultendpunct}{\mcitedefaultseppunct}\relax
\EndOfBibitem
\bibitem[Mejia-Rodriguez \latin{et~al.}(2021)Mejia-Rodriguez, Kunitsa, Apr\`a,
  and Govind]{GWnwchem}
Mejia-Rodriguez,~D.; Kunitsa,~A.~A.; Apr\`a,~E.; Govind,~N. Scalable molecular
  GW calculations: Valence and core spectra. \emph{J. Chem. Theory Comput.}
  \textbf{2021}, \emph{17}, 7504--7517\relax
\mciteBstWouldAddEndPuncttrue
\mciteSetBstMidEndSepPunct{\mcitedefaultmidpunct}
{\mcitedefaultendpunct}{\mcitedefaultseppunct}\relax
\EndOfBibitem
\bibitem[Mejia-Rodriguez \latin{et~al.}(2022)Mejia-Rodriguez, Kunitsa, AprÃ ,
  and Govind]{DMRGWBasis}
Mejia-Rodriguez,~D.; Kunitsa,~A.; AprÃ ,~E.; Govind,~N. Basis Set Selection
  for Molecular Core-Level GW Calculations. \emph{Journal of Chemical Theory
  and Computation} \textbf{2022}, \emph{18}, 4919--4926, PMID: 35816679\relax
\mciteBstWouldAddEndPuncttrue
\mciteSetBstMidEndSepPunct{\mcitedefaultmidpunct}
{\mcitedefaultendpunct}{\mcitedefaultseppunct}\relax
\EndOfBibitem
\bibitem[H{\"a}ttig(2005)]{RIFIT}
H{\"a}ttig,~C. Optimization of auxiliary basis sets for RI-MP2 and RI-CC2
  calculations: Core-valence and quintuple-$\zeta$ basis sets for H to Ar and
  QZVPP basis sets for Li to Kr. \emph{Phys. Chem. Chem. Phys.} \textbf{2005},
  \emph{7}, 59--66\relax
\mciteBstWouldAddEndPuncttrue
\mciteSetBstMidEndSepPunct{\mcitedefaultmidpunct}
{\mcitedefaultendpunct}{\mcitedefaultseppunct}\relax
\EndOfBibitem
\bibitem[Pritchard \latin{et~al.}(2019)Pritchard, Altarawy, Didier, Gibson, and
  Windus]{BSE}
Pritchard,~B.~P.; Altarawy,~D.; Didier,~B.; Gibson,~T.~D.; Windus,~T.~L. A new
  basis set exchange: an open, up-to-date resource for the molecular sciences
  community. \emph{J. Chem. Inf. Model.} \textbf{2019}, \emph{59},
  4814--4820\relax
\mciteBstWouldAddEndPuncttrue
\mciteSetBstMidEndSepPunct{\mcitedefaultmidpunct}
{\mcitedefaultendpunct}{\mcitedefaultseppunct}\relax
\EndOfBibitem
\bibitem[Stoychev \latin{et~al.}(2017)Stoychev, Auer, and
  Neese]{Stoychev:2017:554}
Stoychev,~G.~L.; Auer,~A.~A.; Neese,~F. Automatic generation of auxiliary basis
  sets. \emph{J. Chem. Theory Comput.} \textbf{2017}, \emph{13}, 554--562\relax
\mciteBstWouldAddEndPuncttrue
\mciteSetBstMidEndSepPunct{\mcitedefaultmidpunct}
{\mcitedefaultendpunct}{\mcitedefaultseppunct}\relax
\EndOfBibitem
\bibitem[Fischer \latin{et~al.}(2016)Fischer, Ueltschi, El-Khoury, Mifflin,
  Hess, Wang, Cramer, and Govind]{fischer:2016:1429}
Fischer,~S.~A.; Ueltschi,~T.~W.; El-Khoury,~P.~Z.; Mifflin,~A.~L.; Hess,~W.~P.;
  Wang,~H.-F.; Cramer,~C.~J.; Govind,~N. Infrared and Raman spectroscopy from
  ab initio molecular dynamics and static normal mode analysis: The C--H region
  of DMSO as a case study. \emph{The Journal of Physical Chemistry B}
  \textbf{2016}, \emph{120}, 1429--1436\relax
\mciteBstWouldAddEndPuncttrue
\mciteSetBstMidEndSepPunct{\mcitedefaultmidpunct}
{\mcitedefaultendpunct}{\mcitedefaultseppunct}\relax
\EndOfBibitem
\bibitem[Nos\'e(1984)]{Nose:1984:511}
Nos\'e,~S. A unified formulation of constant temperature molecular dynamics
  methods. \emph{J. Chem. Phys.} \textbf{1984}, \emph{81}, 511--519\relax
\mciteBstWouldAddEndPuncttrue
\mciteSetBstMidEndSepPunct{\mcitedefaultmidpunct}
{\mcitedefaultendpunct}{\mcitedefaultseppunct}\relax
\EndOfBibitem
\bibitem[Hoover(1985)]{Hoover:1985:1695}
Hoover,~W.~G. Canonical dynamics: Equilibrium phase-space distributions.
  \emph{Phys. Rev. A} \textbf{1985}, \emph{31}, 1695\relax
\mciteBstWouldAddEndPuncttrue
\mciteSetBstMidEndSepPunct{\mcitedefaultmidpunct}
{\mcitedefaultendpunct}{\mcitedefaultseppunct}\relax
\EndOfBibitem
\bibitem[Martyna \latin{et~al.}(1992)Martyna, Klein, and
  Tuckerman]{Martyna:1992:2635}
Martyna,~G.~J.; Klein,~M.~L.; Tuckerman,~M. Nos\'e-Hoover chains: The canonical
  ensemble via continous dynamics. \emph{J. Chem. Phys.} \textbf{1992},
  \emph{97}, 2635--2643\relax
\mciteBstWouldAddEndPuncttrue
\mciteSetBstMidEndSepPunct{\mcitedefaultmidpunct}
{\mcitedefaultendpunct}{\mcitedefaultseppunct}\relax
\EndOfBibitem
\bibitem[Weigend(2006)]{JFIT}
Weigend,~F. Accurate Coulomb-fitting basis sets for H to Rn. \emph{Phys. Chem.
  Chem. Phys.} \textbf{2006}, \emph{8}, 1005--1065\relax
\mciteBstWouldAddEndPuncttrue
\mciteSetBstMidEndSepPunct{\mcitedefaultmidpunct}
{\mcitedefaultendpunct}{\mcitedefaultseppunct}\relax
\EndOfBibitem
\bibitem[Miyanaga \latin{et~al.}(1988)Miyanaga, Watanabe, and
  Ikeda]{Miyanaga:1988}
Miyanaga,~T.; Watanabe,~I.; Ikeda,~S. Amplitude in EXAFS and Ligand Exchange
  Reaction of Hydrated 3d Transition Metal Complexes. \emph{Chem. Lett.}
  \textbf{1988}, \emph{17}, 1073--1076\relax
\mciteBstWouldAddEndPuncttrue
\mciteSetBstMidEndSepPunct{\mcitedefaultmidpunct}
{\mcitedefaultendpunct}{\mcitedefaultseppunct}\relax
\EndOfBibitem
\bibitem[Lundberg \latin{et~al.}(2007)Lundberg, Ullstr{\"o}m, D'Angelo, and
  Persson]{Lundberg:2007}
Lundberg,~D.; Ullstr{\"o}m,~A.-S.; D'Angelo,~P.; Persson,~I. A structural study
  of the hydrated and the dimethylsulfoxide, $N$,$N'$-dimethylpropyleneurea,
  and $N$,$N$-dimethylthioformamide solvated iron(II) and iron(III) ions in
  solution and solid state. \emph{Inorg. Chim. Acta} \textbf{2007}, \emph{360},
  1809--1818\relax
\mciteBstWouldAddEndPuncttrue
\mciteSetBstMidEndSepPunct{\mcitedefaultmidpunct}
{\mcitedefaultendpunct}{\mcitedefaultseppunct}\relax
\EndOfBibitem
\bibitem[Fulton \latin{et~al.}(2012)Fulton, Bylaska, Bogatko, Balasubramanian,
  Cau{\"e}t, Schenter, and Weare]{Fulton:2012:2588}
Fulton,~J.~L.; Bylaska,~E.~J.; Bogatko,~S.; Balasubramanian,~M.; Cau{\"e}t,~E.;
  Schenter,~G.~K.; Weare,~J.~H. Near-quantitative agreement of model-free
  DFT-MD predictions with XAFS observations of the hydration structure of
  highly charged transition-metal ions. \emph{J. Phys. Chem. Lett.}
  \textbf{2012}, \emph{3}, 2588--2593\relax
\mciteBstWouldAddEndPuncttrue
\mciteSetBstMidEndSepPunct{\mcitedefaultmidpunct}
{\mcitedefaultendpunct}{\mcitedefaultseppunct}\relax
\EndOfBibitem
\bibitem[Persson \latin{et~al.}(2020)Persson, Lundberg, Bajn{\'o}czi,
  Klementiev, Just, and Clauss]{Persson:2020:9538}
Persson,~I.; Lundberg,~D.; Bajn{\'o}czi,~{\'E}.~G.; Klementiev,~K.; Just,~J.;
  Clauss,~K. G. V.~S. EXAFS study on the coordination chemistry of the solvated
  copper(II) ion in a series of oxygen donor solvents. \emph{Inorg. Chem.}
  \textbf{2020}, \emph{59}, 9538--9550\relax
\mciteBstWouldAddEndPuncttrue
\mciteSetBstMidEndSepPunct{\mcitedefaultmidpunct}
{\mcitedefaultendpunct}{\mcitedefaultseppunct}\relax
\EndOfBibitem
\bibitem[Krishnan \latin{et~al.}(1980)Krishnan, Binkley, Seeger, and
  Pople]{krishnan1980a}
Krishnan,~R.; Binkley,~J.~S.; Seeger,~R.; Pople,~J.~A. Self-consistent
  molecular orbital methods. XX. A basis set for correlated wave functions.
  \emph{J. Chem. Phys.} \textbf{1980}, \emph{72}, 650--654\relax
\mciteBstWouldAddEndPuncttrue
\mciteSetBstMidEndSepPunct{\mcitedefaultmidpunct}
{\mcitedefaultendpunct}{\mcitedefaultseppunct}\relax
\EndOfBibitem
\bibitem[Duchemin \latin{et~al.}(2016)Duchemin, Jacquemin, and
  Blase]{Duchemin:2016:164106}
Duchemin,~I.; Jacquemin,~D.; Blase,~X. Combining the $GW$ formalism with the
  polarizable continuum model: A state-specific non-equilibrium approach.
  \emph{J. Chem. Phys.} \textbf{2016}, \emph{144}, 164106\relax
\mciteBstWouldAddEndPuncttrue
\mciteSetBstMidEndSepPunct{\mcitedefaultmidpunct}
{\mcitedefaultendpunct}{\mcitedefaultseppunct}\relax
\EndOfBibitem
\bibitem[Duan \latin{et~al.}(2022)Duan, Chu, Nandy, and Kulik]{Duan:2022:4962}
Duan,~C.; Chu,~D. B.~K.; Nandy,~A.; Kulik,~H.~J. Detection of multi-reference
  character imbalances enables a transfer learning approach for virtual high
  throughput screening with coupled cluster accuracy at DFT cost. \emph{Chem.
  Sci.} \textbf{2022}, \emph{13}, 4962--4971\relax
\mciteBstWouldAddEndPuncttrue
\mciteSetBstMidEndSepPunct{\mcitedefaultmidpunct}
{\mcitedefaultendpunct}{\mcitedefaultseppunct}\relax
\EndOfBibitem
\bibitem[Lesiuk and Jeziorski(2019)Lesiuk, and Jeziorski]{CBSRiemann}
Lesiuk,~M.; Jeziorski,~B. Complete basis set extrapolation of electronic
  correlation energies using the Riemann zeta function. \emph{J. Chem. Theory
  Comput.} \textbf{2019}, \emph{15}, 5398--5403\relax
\mciteBstWouldAddEndPuncttrue
\mciteSetBstMidEndSepPunct{\mcitedefaultmidpunct}
{\mcitedefaultendpunct}{\mcitedefaultseppunct}\relax
\EndOfBibitem
\bibitem[Hung \latin{et~al.}(2017)Hung, Bruneval, Baishya, and
  \"O\u{g}\"ut]{Hung:2017:2135}
Hung,~L.; Bruneval,~F.; Baishya,~K.; \"O\u{g}\"ut,~S. Benchmarking the GW
  Approximation and Bethe–Salpeter Equation for Groups IB and IIB Atoms and
  Monoxides. \emph{J. Chem. Theory Comput.} \textbf{2017}, \emph{13},
  2135--2146\relax
\mciteBstWouldAddEndPuncttrue
\mciteSetBstMidEndSepPunct{\mcitedefaultmidpunct}
{\mcitedefaultendpunct}{\mcitedefaultseppunct}\relax
\EndOfBibitem
\bibitem[Pasquarello \latin{et~al.}(2001)Pasquarello, Petri, Salmon, Parisel,
  Car, Éva Tóth, Powell, Fischer, Helm, and Merbach]{Pasquarello:2001:856}
Pasquarello,~A.; Petri,~I.; Salmon,~P.~S.; Parisel,~O.; Car,~R.; Éva Tóth,;
  Powell,~D.~H.; Fischer,~H.~E.; Helm,~L.; Merbach,~A.~E. First Solvation Shell
  of the Cu(II) Aqua Ion: Evidence for Fivefold Coordination. \emph{Science}
  \textbf{2001}, \emph{291}, 856--859\relax
\mciteBstWouldAddEndPuncttrue
\mciteSetBstMidEndSepPunct{\mcitedefaultmidpunct}
{\mcitedefaultendpunct}{\mcitedefaultseppunct}\relax
\EndOfBibitem
\bibitem[Frank \latin{et~al.}(2018)Frank, Benfatto, and
  Qayyum]{Frank:2018:204302}
Frank,~P.; Benfatto,~M.; Qayyum,~M. [Cu(aq)]2+ is structurally plastic and the
  axially elongated octahedron goes missing. \emph{J. Chem. Phys.}
  \textbf{2018}, \emph{148}, 204302\relax
\mciteBstWouldAddEndPuncttrue
\mciteSetBstMidEndSepPunct{\mcitedefaultmidpunct}
{\mcitedefaultendpunct}{\mcitedefaultseppunct}\relax
\EndOfBibitem
\bibitem[Helm and Merbach(2005)Helm, and Merbach]{Helm:2005:1923}
Helm,~L.; Merbach,~A.~E. Inorganic and Bioinorganic Solvent Exchange
  Mechanisms. \emph{Chem. Rev.} \textbf{2005}, \emph{105}, 1923--1960\relax
\mciteBstWouldAddEndPuncttrue
\mciteSetBstMidEndSepPunct{\mcitedefaultmidpunct}
{\mcitedefaultendpunct}{\mcitedefaultseppunct}\relax
\EndOfBibitem
\bibitem[Lincoln(2005)]{Lincoln:2005:523}
Lincoln,~S.~F. Mechanistic Studies of Metal Aqua Ions: A Semi-Historical
  Perspective. \emph{Helvetica Chimica Acta} \textbf{2005}, \emph{88},
  523--545\relax
\mciteBstWouldAddEndPuncttrue
\mciteSetBstMidEndSepPunct{\mcitedefaultmidpunct}
{\mcitedefaultendpunct}{\mcitedefaultseppunct}\relax
\EndOfBibitem
\bibitem[Wertheim \latin{et~al.}(1974)Wertheim, Butler, West, and
  Buchanan]{Wertheim:1974:1369}
Wertheim,~G.~K.; Butler,~M.~A.; West,~K.~W.; Buchanan,~D. N.~E. Determination
  of the Gaussian and Lorentzian content of experimental line shapes.
  \emph{Rev. Sci. Instrum.} \textbf{1974}, \emph{45}, 1369--1371\relax
\mciteBstWouldAddEndPuncttrue
\mciteSetBstMidEndSepPunct{\mcitedefaultmidpunct}
{\mcitedefaultendpunct}{\mcitedefaultseppunct}\relax
\EndOfBibitem
\bibitem[Knizia(2013)]{Knizia:2013}
Knizia,~G. Intrinsic Atomic Orbitals: An Unbiased Bridge between Quantum Theory
  and Chemical Concepts. \emph{J. Chem. Theory Comput.} \textbf{2013},
  \emph{9}, 4834--4843\relax
\mciteBstWouldAddEndPuncttrue
\mciteSetBstMidEndSepPunct{\mcitedefaultmidpunct}
{\mcitedefaultendpunct}{\mcitedefaultseppunct}\relax
\EndOfBibitem
\end{mcitethebibliography}

%\section{Supporting Information}
%\input{supporting.tex}

\end{document}

\maketitle

\section{Updates to COSMO parameters}

The original COSMO parameters in NWChem lead to a situation where the
solvation shells are effectively screened from one another by a set of
COSMO charges in between them.

This situation can be remedied by using atomic radii, $R_{at}$ 
which are almost twice as large as the standard ones. 
The radii mostly used in the literature are defined as $1.1 \times$ 
the Bondi radii, whereas $2\times$ the Bondi radii are needed 
to eliminate most of the spurious COSMO charges.

An alternative comes from redefining the York-Karplus switching
function parameters.
\begin{equation}
    S_{YK}(r) = \begin{cases} 
    0 & r < 0 \\
    r^3 ( 10 -15r + 6r^2) & 0 \leqslant r \leqslant 1 \\
    1 & r > 1
    \end{cases}
\end{equation}
For a point $k$ located at $\mathbf{r}_{ik} = \mathbf{R}_i + R_{at,i}\hat{\mathbf{r}}_k$ and belonging to atom $i$, 
the switching function assigns a weight depending on the position of all other atoms
\begin{equation}
    S_{ik} = \prod\limits_{j\neq i} S_{YK}\left(r_{ik,j}\right)
\end{equation}
\begin{equation}
    r_{ik,j} = \frac{\lvert \mathbf{r}_{ik} - \mathbf{R}_j \rvert - R_{in,j}}{R_{out,j} - R_{in,j}}
\end{equation}
Typically, the inner and outer atomic radius are given as fractions
of the atomic radius
\begin{equation}
    R_{in,i} = \left( 1 - \alpha\gamma \right) R_{at,i}
\end{equation}
\begin{equation}
    R_{out,i} = \left( 1 + (1-\alpha)\gamma\right) R_{at,i}
\end{equation}
In \textsc{NWChem}, $\alpha = 0.5$ and 
$\gamma = \sqrt{0.25^{\text{minbem}}}$, with $\text{minbem}=2$ as
default. This leads to define the inner and outer radii as
\begin{equation}
 R_{in,i} = 0.875 R_{at,i}   
\end{equation}
\begin{equation}
    R_{out,i} = 1.125 R_{at,i}
\end{equation}

The following figure shows a schematic representation
of COSMO surface built from two atoms. The dots represent
the position of some COSMO charges and are located at the
$R_{at,i}$ distance from each atomic center. The red arcs
are at $R_{in,i}$ distance, while the blue arcs are at
$R_{out,i}$ distance. All the points penetrating the
inner arcs into an atomic center get discarded by the
switching function, but all other points remain. With
the standard \textsc{NWChem} parameters, many points inside
the dotted arcs will survive the pruning process. However,
this points are not located at the surface and, depending
on the exact geometry, can effectively screen one atom from
another even at distances smaller than the sum of the
respective atomic radii.

In order to alleviate this issue, one can set $\alpha = 0$ to get
\begin{equation}
    R_{in,i} = R_{at,i}
\end{equation}
\begin{equation}
    R_{out,i} = \left(1 + \gamma\right)R_{at,i}
\end{equation}
In this case, COSMO charges will only start to appear in
between atoms only when its distance is larger than the
sum of their atomic radii.

\begin{figure}
\begin{center}
\begin{tikzpicture}
\draw[very thick, dotted] (0.8281,0.5605) arc (34.09:325.91:1.0);
\draw[thick, red] (0.75,0.4507) arc (31:329:0.875);
\draw[thick, blue] (0.9167,0.6522) arc (35.43:324.57:1.125);
\draw[very thick, dotted, green!50!black] (0.6719,0.5605) arc (145.90: -145.90: 1.0);
\draw[thick, red] (0.75,0.4507) arc (149:-149:0.875);
\draw[thick, blue] (0.5833,0.6522) arc (144.57:-144.57:1.125);
\end{tikzpicture}
\caption{Schematic representation of a COSMO surface.}
\end{center}
\end{figure}

Further reductions in the number of points appearing in between
atoms interacting non-covalently can be obtained by slightly
increasing the definition of the atomic radii. We found that
increasing the Bondi radii by 22\%, instead of the typical
20\%, provides very good results for water clusters.

An actual comparison between the old and new switching function
parameters is shown in Figure \ref{fig:COSMO} for a snapshot of
the \ce{[Fe(H2O)6]^{3+}(H2O)54} cluster. The COSMO charges were
generated using a Fibonacci spiral grid with 301 points per
atom, with a further minimization of the Coulomb repulsion 
between the point charges. After applying the corresponding
switching functions, the remaining number of COSMO charges
were 23085 using the default parameters, and 8944 using the
modified parameters. Although the top images of Figure \ref{fig:COSMO}
show that many of the removed points were on the surface
(the orange spheres make a denser surface than the green ones),
many points were also removed from the inside. The bottom images
of Figure \ref{fig:COSMO} show that the default
switching function parameters lead to many COSMO charges appearing
inside the surface in between each solvation shell, while the
modified parameters offer a nice hollow cavity.

Structurally, the overall result of using the modified parameters 
is that the metal-oxygen covalent bonds get slightly longer, 
while the distance between each water shell gets slightly shorter. 
The eigenvalue spectrum gets modified as well, resulting in slightly 
deeper occupied orbitals because the charge at the metal ion is not overscreened. 

\begin{figure}
    \centering
    \fbox{\includegraphics[trim={275 220 220 250},clip,width=0.45\textwidth]{figures/oldsurf}}
    \fbox{\includegraphics[trim=275 220 220 250,clip, width=0.45\textwidth]{figures/newsurf}} \\
    \fbox{\includegraphics[width=0.4\textwidth]{figures/oldsurf_inside_noao}}
    \fbox{\includegraphics[width=0.4\textwidth]{figures/newsurf_inside_noao}}
    \caption{Comparison between the COSMO surfaces generated for \ce{[Fe(H2O)6]^{3+}(H2O)54} with default parameters (orange spheres) and with the modifications proposed (green spheres). The top images show general views of the surfaces, while the bottom images show a view from the outermost oxygen. }
    \label{fig:COSMO}
\end{figure}

\section{Tables}

\begin{table}[]
    \renewcommand{\arraystretch}{1.1}
    \centering
    \caption{Metal-Oxygen bond distances [\AA] obtained with the def2-tzvp/sapporo-tzp-2012 basis set combination for the \ce{[M(H2O)40]^{n+}} clusters}
    \label{tab:my_label}
    \begin{tabular}{c c c c c c c c}
                                            &           & O1 & O2 & O3 & O4 & O5 & O6 \\\hline
    \multirow{3}{*}{\ce{[V(H2O)40]^{3+}}}   & PBE       & 1.951 & 2.016 & 2.040 & 2.089 & 2.101 & 1.967   \\
                                            & PBE0      & 1.948 & 1.991 & 2.020 & 2.067 & 2.066 & 1.966 \\
                                            & r$^2$SCAN & 1.945 & 1.997 & 2.026 & 2.076 & 2.069 & 1.963 \\\hline
    \multirow{3}{*}{\ce{[Cr(H2O)40]^{3+}}}  & PBE      & 1.975 & 2.020 & 1.969 & 2.026 & 2.024 & 1.975   \\
                                            & PBE0      & 1.957 & 1.993 & 1.954 & 2.000 & 1.997 & 1.962 \\
                                            & r$^2$SCAN & 1.967 & 2.000 & 1.960 & 2.001 & 1.994 & 1.977 \\\hline
    \multirow{3}{*}{\ce{[Cr(H2O)40]^{2+}}}  & PBE       & 2.050 & 2.050 & 2.032 & 2.056 & 2.524 & 2.409 \\
                                            & PBE0      & 2.047 & 2.074 & 2.047 & 2.074 & 2.476 & 2.348 \\
                                            & r$^2$SCAN & 2.051 & 2.064 & 2.034 & 2.063 & 2.498 & 2.346 \\\hline
    \multirow{3}{*}{\ce{[Mn(H2O)40]^{2+}}}  & PBE       & 2.150 & 2.176 & 2.139 & 2.203 & 2.298 & 2.257 \\
                                            & PBE0      & 2.149 & 2.166 & 2.140 & 2.193 & 2.265 & 2.226 \\
                                            & r$^2$SCAN & 2.147 & 2.170 & 2.134 & 2.184 & 2.248 & 2.213 \\\hline
    \multirow{3}{*}{\ce{[Fe(H2O)40]^{3+}}}  & PBE       & 2.012 & 2.038 & 1.957 & 2.077 & 2.116 & 2.072  \\
                                            & PBE0      & 1.988 & 2.011 & 1.962 & 2.043 & 2.067 & 2.009 \\
                                            & r$^2$SCAN & 2.001 & 2.025 & 1.948 & 2.050 & 2.063 & 2.029 \\\hline
    \multirow{3}{*}{\ce{[Fe(H2O)40]^{2+}}}   & PBE       & 2.075 & 2.105 & 2.098 & 2.164 & 2.278 & 2.184 \\
                                            & PBE0      & 2.083 & 2.113 & 2.140 & 2.196 & 2.171 & 2.155 \\
                                            & r$^2$SCAN & 2.068 & 2.097 & 2.096 & 2.147 & 2.210 & 2.143 \\\hline
    \multirow{3}{*}{\ce{[Co(H2O)40]^{2+}}}   & PBE       & 2.053 & 2.149 & 2.057 & 2.195 & 2.114 & 2.181 \\
                                            & PBE0      & 2.064 & 2.137 & 2.052 & 2.149 & 2.111 & 2.142 \\
                                            & r$^2$SCAN & 2.048 & 2.129 & 2.045 & 2.146 & 2.093 & 2.134 \\\hline
    \multirow{3}{*}{\ce{[Ni(H2O)40]^{2+}}}   & PBE       & 2.156 & 2.119 & 2.046 & 2.080 & 2.100 & 2.043 \\
                                            & PBE0      & 2.141 & 2.096 & 2.034 & 2.081 & 2.028 & 2.116 \\
                                            & r$^2$SCAN & 2.129 & 2.090 & 2.034 & 2.078 & 2.028 & 2.111 \\\hline
    \multirow{3}{*}{\ce{[Cu(H2O)40]^{2+}}}   & PBE       & 1.953 & 1.954 & 1.982 & 2.025 & 2.499 & 3.068   \\
                                            & PBE0      & 1.947 & 1.937 & 1.965 & 2.001 & 2.440 & 2.700  \\
                                            & r$^2$SCAN & 1.948 & 1.948 & 1.971 & 2.006 & 2.466 & 2.529   \\\hline
    \end{tabular}
\end{table}

\begin{table}[]
\renewcommand{\arraystretch}{1.2}
\centering
\caption{$G_0W_0$@DFT vertical ionization potentials, in eV, obtained with the basis set combination def2-tzvp/sapporo-tzp-2012 as described in the main text. All experimental values were obtained from Yepes and cols.\cite{Yepes:2014:6850}}
\begin{tabular}{c c c c c c c c c c c c c c c}
    &  Exptl.  & & \multicolumn{3}{c}{PBE} & & \multicolumn{3}{c}{PBE0}  & & \multicolumn{4}{c}{r$^2$SCAN}      \\\cline{4-6}\cline{8-10}\cline{12-15}
             &          & & 6w   & 18w & 40w  & & 6w & 18w & 40w & & 6w   & 18w & 40w  & 60w\\\hline
 \ce{V^3+}   &   8.41   & & 9.36 & 7.55& 6.87 & &11.16& 9.41& 8.79& & 9.99 & 8.19 &7.57  &  7.65 \\     
 \ce{Cr^3+}  &   9.48   & & 10.66& 8.38& 8.04 & &12.48&10.31&10.09& &11.38 &9.09  &8.81  &  8.86 \\
 \ce{Cr^2+}  &   6.76   & &  6.85& 5.27& 5.54 & & 7.77& 6.75& 6.45& & 6.80 & 5.77  &5.36 &  5.54 \\
 \ce{Mn^2+}  &   8.82   & & 9.03 & 7.81& 7.32 & &10.33& 9.14&8.79 & & 9.42 &8.21  &7.83  &  8.04 \\
 \ce{Fe^3+}  &  10.28   & & 11.61& 9.46& 9.09 & &13.88&11.67&11.20& &12.54 &10.38 &10.10 &  9.96 \\
 \ce{Fe^2+}  &   7.13   & &  6.91& 5.69& 5.35 & & 8.13& 7.00& 6.82& & 7.54 & 6.36 & 5.74 &  5.87 \\
 \ce{Co^2+}  &   8.70   & & 8.59 & 7.17& 6.92 & &10.28& 9.05& 8.99& & 9.07 &7.78  &7.63  &  7.44 \\
 \ce{Ni^2+}  &   9.45   & & 9.19 & 7.82& 7.87 & &11.40&10.12&10.20& &10.11 &8.80  &8.76  &  8.92 \\
 \ce{Cu^2+}  &   9.65   & & 9.70 & 8.55& 8.09 & &11.36&10.28&10.28& & 9.82 &8.75  &8.74  &  8.95 \\\hline
\end{tabular}
\label{tab:my_label}
\end{table}

\begin{table}[]
\renewcommand{\arraystretch}{1.2}
\centering
\caption{Average metal-oxygen bond distances [\AA] from theory and experiment. }
\begin{tabular}{c c c c c c c c c c c c c c  c }
    &  Exptl.  & & \multicolumn{3}{c}{PBE} & & \multicolumn{3}{c}{PBE0}  & & \multicolumn{4}{c}{r$^2$SCAN}      \\\cline{4-6}\cline{8-10}\cline{12-15}
             &       & & 6w & 18w   & 40w   & & 6w   & 18w  & 40w & & 6w   & 18w & 40w  & 60w\\\hline 
\ce{V^3+}    & 2.000 & & 2.009 & 2.017 & 2.027 & & 1.998 & 2.002 & 2.010 & & 1.999 & 2.003 & 2.013 & 2.011 \\
\ce{Cr^3+}   & 1.972 & & 1.994 & 1.994 & 1.998 & & 1.965 & 1.974 & 1.977 & & 1.967 & 1.977 & 1.983 & 1.981\\
\ce{Cr^2+}   & 2.080 & & 2.073 & 2.064 & 2.047 & & 2.071 & 2.067 & 2.060 & & 2.068 & 2.062 & 2.053 & 2.053 \\
\ce{Mn^2+}   & 2.172 & & 2.187 & 2.200 & 2.204 & & 2.175 & 2.186 & 2.190 & & 2.166 & 2.177 & 2.183 & 2.180 \\
\ce{Fe^3+}   & 2.006 & & 2.019 & 2.033 & 2.045 & & 1.997 & 2.007 & 2.013 & & 1.998 & 2.009 & 2.019 & 2.016 \\
\ce{Fe^2+}   & 2.118 & & 2.138 & 2.150 & 2.151 & & 2.126 & 2.136 & 2.143 & & 2.114 & 2.123 & 2.128 & 2.134 \\
\ce{Co^2+}   & 2.080 & & 2.102 & 2.107 & 2.125 & & 2.089 & 2.093 & 2.109 & & 2.081 & 2.084 & 2.099 & 2.092 \\
\ce{Ni^2+}   & 2.051 & & 2.079 & 2.087 & 2.101 & & 2.075 & 2.078 & 2.083 & & 2.049 & 2.075 & 2.078 & 2.077 \\
\ce{Cu^2+}   & 1.957 & & 2.019 & 2.010 & 1.979 & & 1.998 & 1.992 & 1.963 & & 1.994 & 1.989 & 1.968 & 1.989 \\\hline
\end{tabular}
\label{tab:my_label2}
\end{table}

\newpage
\section{Figures}

\begin{figure}
    \centering
    \includegraphics[width=0.85\textwidth]{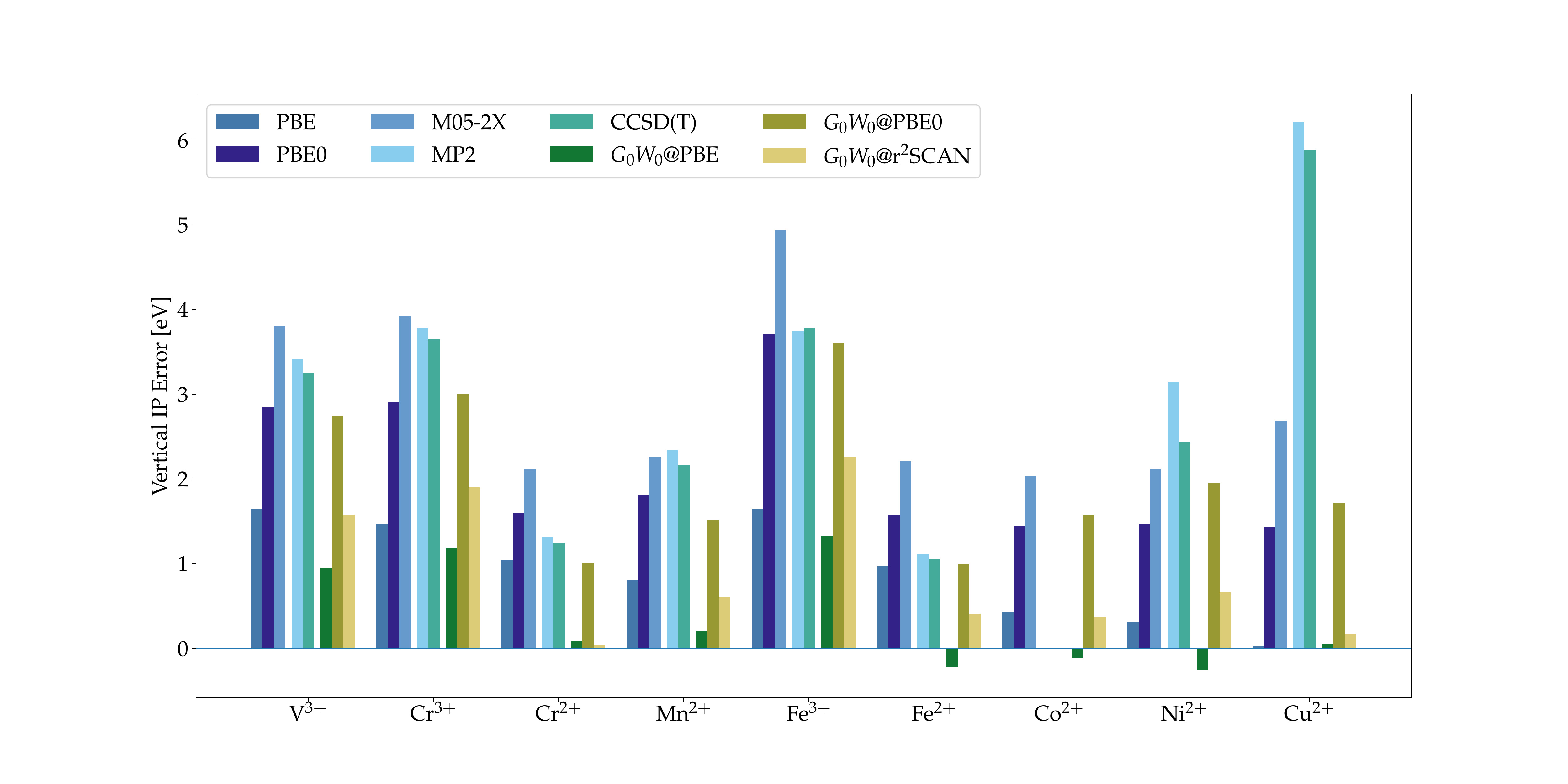}\\
    \includegraphics[width=0.85\textwidth]{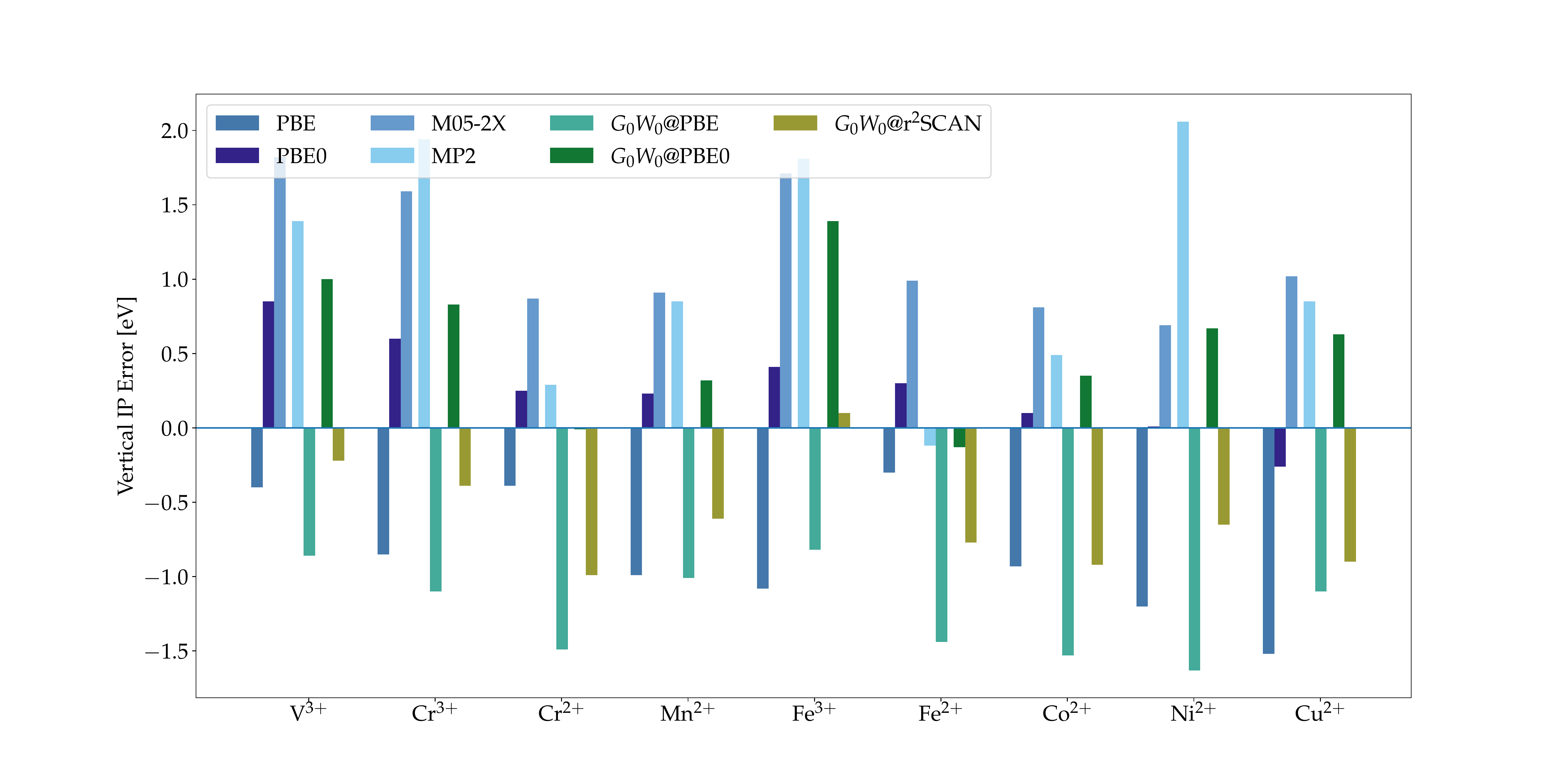}
    \caption{Vertical ionization potential errors [eV] with respect to experiment for the 6-water clusters (top), and 18-water clusters (bottom).The $\Delta$DFT, $\Delta$MP2, and $\Delta$CCSD(T) values were taken from Ref. \citenum{Yepes:2014:6850}. $G_0W_0$ values were obtained with the triple-$\zeta$ combination.}
\end{figure}

\newpage
\begin{figure}
    \centering
    \includegraphics[width=0.85\textwidth]{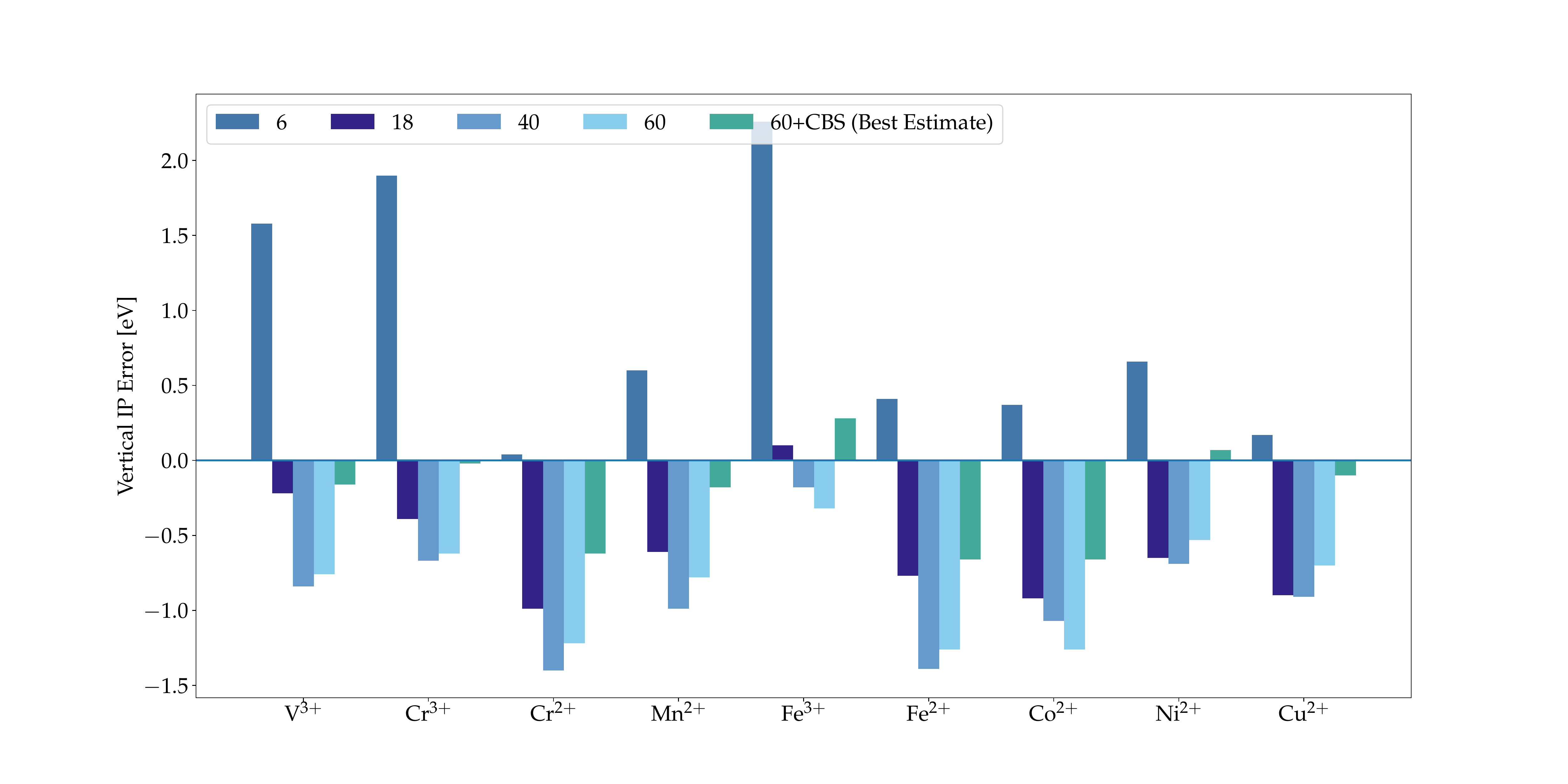}
    \caption{Vertical ionization potential errors [eV] with respect to experiment for the $G_0W_0$@r$^2$SCAN approach. All values, except for the best estimate, were obtained at the triple-$\zeta$ level. The best estimate value includes the complete basis set extrapolation shift found using the Riemann-$\zeta(3)$ method.}
\end{figure}

\bibliography{bibliography.bib,gw.bib}